\documentclass[final,5p,times,twocolumn,square,sort,comma,numbers]{elsarticle}

\usepackage{xurl}
\usepackage{amsthm}

\usepackage{amssymb}
\usepackage{multicol}
\usepackage{xcolor}

\usepackage[normalem]{ulem}

\usepackage{booktabs}
\usepackage[labelfont=bf]{caption}
\usepackage{longtable}
\usepackage{hyperref}
\usepackage{array}

\usepackage{placeins}

\usepackage{subcaption}

 \usepackage{amsmath} 
\usepackage{nomencl} 
\usepackage{framed} 
\usepackage{enumitem}
\usepackage{multirow}
\usepackage{tikz}
\usepackage{float}
\usetikzlibrary{arrows.meta, positioning, backgrounds}

\setlist{noitemsep, topsep=2pt, partopsep=0pt}
\definecolor{refblue}{RGB}{33,113,181}
\definecolor{predorange}{RGB}{230,85,13}
\definecolor{tubegray}{gray}{0.85}
\definecolor{horizon}{RGB}{102,194,165}
\makenomenclature

\renewcommand*\nompreamble{\begin{multicols}{2}}

\renewcommand*\nompostamble{\end{multicols}}
\newcolumntype{C}[1]{>{\centering\arraybackslash}p{#1}}

\usepackage{etoolbox}
\renewcommand\nomgroup[1]{%
  \item[\bfseries
   \ifstrequal{#1}{A}{Acronyms}{%
  \ifstrequal{#1}{I}{Indices}{%
  \ifstrequal{#1}{V}{Variables}{}}}%
]}
\definecolor{cardinal} {RGB}{196, 30, 58}
\definecolor{lightgrey}{RGB}{150,150,150}

\journal{Journal of Process Control}
\begin{document}

\begin{frontmatter}



\title{Grid-Compatible Flexibility from Multi-Energy Systems via Cyclic-Terminal Economic MPC with Hybrid Thermal–Electrical Dynamics}


\author[abu,ditens]{Abdul Azzam \corref{cor1}}
\affiliation[abu]{organization={University of Stuttgart, Institute of Energy Economics and Rational Energy Use},
            addressline={Heßbrühlstrasse 49a},
            city={Stuttgart},
            postcode={70565},
            state={Baden-Württemberg},
            country={Germany}}

\affiliation[ist]{organization={University of Stuttgart, Institute for Systems Theory and Automatic Control},%
            addressline={Pfaffenwaldring 9},
            city={Stuttgart},
            postcode={70569},
            state={Baden-Württemberg},
            country={Germany}}
\affiliation[ditens]{organization={Stuttgart Research Initiative "Discursive Transformation of Energy Systems" (SRI DiTEnS)
},
            addressline={Heßbrühlstrasse 49a},
            city={Stuttgart},
            postcode={70565},
            state={Baden-Württemberg},
            country={Germany}}

\ead{abdulrahmman.azzam@ier.uni-stuttgart.de}
\cortext[cor1]{Corresponding author}

\author[ist]{Lukas Schwenkel}
\author[abu,ditens]{Leon Scheurer}
\author[abu,ditens]{Pascal Häbig}
\author[abu,ditens]{Kai Hufendiek}

\begin{abstract}
Coupled electrical and thermal infrastructures need controllers
that respond to market prices and still solve fast enough to run
online. This paper presents a unified Economic Model Predictive
Control (EMPC) framework for the coordinated operation of
integrated thermal and electrical energy networks. Building on
cyclic-terminal EMPC, the proposed approach incorporates hybrid
thermal--electrical dynamics, network constraints, and time-varying
economic signals within a single mixed-integer state-space
representation, jointly optimizing combined heat and power units,
large-scale heat pumps, thermal energy storage, batteries, and
grid interactions under a convex economic stage cost.
Computational tractability is ensured by reduced-order models of
district heating networks and DC power flow grids.

The framework is demonstrated on a campus-scale multi-energy
system under time-varying prices and demand profiles.
A joint sweep of the prediction horizon against the terminal
penalty weight shows that the two act as substitutes rather than as
independent tuning knobs. Without terminal anchoring, the closed-loop
cost approaches the periodic-reference average-performance bound only
once the horizon spans several diurnal cycles. With a sufficiently
large terminal weight, the bound is attained essentially tightly at
every tested horizon, including the shortest one, so the horizon
ceases to be a performance-critical parameter and becomes a purely
computational one. The result reproduces on a second, independent
price week. Beyond the weight at which the soft terminal constraint
activates, closed-loop behavior is insensitive to the weight over a
wide multi-decade plateau; below activation, cost and storage tracking
both degrade markedly. These findings are specific to this campus
system and the studied price weeks. A residual receding-horizon drift of the
cost-neutral thermal-storage state is also documented and
interpreted.
\end{abstract}

\end{frontmatter}

\nomenclature[A]{EMPC}{Economic Model Predictive Control}
\nomenclature[A]{MPC}{Model Predictive Control}
\nomenclature[A]{CHP}{Combined Heat and Power}
\nomenclature[A]{HP}{Heat Pump}
\nomenclature[A]{TES}{Thermal Energy Storage}
\nomenclature[A]{DHN}{District Heating Network}
\nomenclature[A]{SOC}{State of Charge}
\nomenclature[A]{PCC}{Point of Common Coupling}
\nomenclature[A]{PV}{Photovoltaic}
\nomenclature[A]{DC}{Direct Current}

\nomenclature[I]{$t$}{Continuous-time variable}
\nomenclature[I]{$k$}{Discrete-time step index}
\nomenclature[I]{$N$}{Prediction horizon length}
\nomenclature[I]{$N_p$}{Period length (number of steps)}
\nomenclature[I]{$T_s$}{Sampling time}
\nomenclature[I]{$\mathcal{N}_e$}{Set of electrical network nodes}
\nomenclature[I]{$\mathcal{E}_e$}{Set of electrical network edges}
\nomenclature[I]{$\mathcal{G}_e$}{Electrical network graph $(\mathcal{N}_e,\mathcal{E}_e,\mathcal{W}_e)$}

\nomenclature[V]{$\mathbf{x}$}{State vector}
\nomenclature[V]{$\mathbf{u}$}{Input (control) vector}
\nomenclature[V]{$\mathbf{d}$}{Disturbance vector}
\nomenclature[V]{$A,B,E$}{State, input, and disturbance matrices}
\nomenclature[V]{$w$}{Affine disturbance vector}
\nomenclature[V]{$\mathcal{X}$}{Admissible state set}
\nomenclature[V]{$\mathcal{U}$}{Admissible input set}
\nomenclature[V]{$\ell(\cdot)$}{Stage cost}
\nomenclature[V]{$J_N$}{Finite-horizon cost}
\nomenclature[V]{$J_f$}{Terminal cost}
\nomenclature[V]{$x^+$}{Successor state in discrete time}
\nomenclature[V]{$x_{\text{ref}}$}{Economic reference state (periodic)}
\nomenclature[V]{$u_{\text{ref}}$}{Economic reference input (periodic)}

\nomenclature[V]{$Q_{\text{TES}}$}{Stored thermal energy in TES [kWh]}
\nomenclature[V]{$Q_{\text{TES,N}}$}{Nominal TES capacity [kWh]}
\nomenclature[V]{$\dot{Q}_{\text{HS}}$}{TES charge/discharge power [kW]}
\nomenclature[V]{$\dot{Q}_{\text{loss}}$}{TES heat losses [kW]}
\nomenclature[V]{$D_{\text{TES}}$}{TES tank diameter [m]}
\nomenclature[V]{$H_{\text{TES}}$}{TES tank height [m]}
\nomenclature[V]{$U_{\text{TES}}$}{TES overall heat-transfer coefficient [W/(m$^2$K)]}
\nomenclature[V]{$S_{\text{TES}}$}{TES external surface area [m$^2$]}
\nomenclature[V]{$T_H,T_L$}{Upper and lower TES temperatures [$^\circ$C]}
\nomenclature[V]{$T_A$}{Ambient air temperature [$^\circ$C]}
\nomenclature[V]{$\dot{Q}_{\max}$}{Maximum TES charge/discharge power [kW]}

\nomenclature[V]{$x_{\text{batt}}$}{Battery state of charge (SOC) [p.u.]}
\nomenclature[V]{$P_{\text{ch}}$}{Battery charging power [kW]}
\nomenclature[V]{$P_{\text{disch}}$}{Battery discharging power [kW]}
\nomenclature[V]{$C_{\text{batt}}$}{Battery capacity [kWh]}
\nomenclature[V]{$\eta_{\text{ch}}$}{Charging efficiency [-]}
\nomenclature[V]{$\eta_{\text{disch}}$}{Discharging efficiency [-]}
\nomenclature[V]{$\gamma_{\text{batt}}$}{Battery mode binary variable (charge/discharge)}
\nomenclature[V]{$\text{SOC}_{\min},\text{SOC}_{\max}$}{SOC limits [p.u.]}
\nomenclature[V]{$P_{\max}$}{Maximum battery power [kW]}

\nomenclature[V]{$P_{\text{HP,el}}$}{Electrical power of heat pump [kW]}
\nomenclature[V]{$P_{\text{HP,th}}$}{Thermal power of heat pump [kW]}
\nomenclature[V]{$\text{COP}$}{Coefficient of performance of the heat pump [-]}
\nomenclature[V]{$\text{COP}_{\text{ref}}$}{Reference COP at nominal temperature lift [-]}
\nomenclature[V]{$\Delta T_{\text{lift}}$}{Temperature lift between source and sink [K]}
\nomenclature[V]{$T_{\text{sink}}$}{Heat sink temperature [$^\circ$C]}
\nomenclature[V]{$T_{\text{source}}$}{Heat source temperature [$^\circ$C]}
\nomenclature[V]{$P_{\text{stat}}$}{Stationary electrical power of heat pump [kW]}

\nomenclature[V]{$P_{\text{set}}$}{CHP setpoint electrical power [kW]}
\nomenclature[V]{$\delta^{\text{on}}$}{CHP on/off state binary variable}
\nomenclature[V]{$\Delta P$}{CHP power gradient command [kW]}
\nomenclature[V]{$\delta^{\text{on,cmd}}$}{CHP on-command binary variable}
\nomenclature[V]{$\delta^{\text{start}}$}{CHP startup indicator binary variable}
\nomenclature[V]{$P_{\text{av}}$}{Available CHP electrical power [kW]}
\nomenclature[V]{$Q_{\text{chp}}$}{CHP thermal output power [kW]}
\nomenclature[V]{$\eta_{\text{el}},\eta_{\text{th}}$}{Electrical and thermal efficiencies of CHP [-]}
\nomenclature[V]{$P_{\min},P_{\max}$}{CHP minimum and maximum power [kW]}

\nomenclature[V]{$\dot{Q}_{\text{coll}}$}{Thermal power from solar collector [kW]}
\nomenclature[V]{$A_{\text{coll}}$}{Collector area [m$^2$]}
\nomenclature[V]{$G_{\text{coll}}$}{Incident solar irradiance on collector [W/m$^2$]}
\nomenclature[V]{$\eta_0$}{Optical efficiency of solar collector [-]}
\nomenclature[V]{$a_1$}{Collector heat loss coefficient [W/(m$^2$K)]}
\nomenclature[V]{$T_{\text{coll,avg}}$}{Mean collector temperature [$^\circ$C]}
\nomenclature[V]{$Q_{\text{load}}$}{Thermal load demand [kW]}
\nomenclature[V]{$P_{\text{load}}$}{Electrical load demand [kW]}

\nomenclature[V]{$T_{\text{DHN}}$}{Average DHN water temperature [$^\circ$C]}
\nomenclature[V]{$T_{\text{supply}}$}{DHN supply temperature [$^\circ$C]}
\nomenclature[V]{$T_{\text{return}}$}{DHN return temperature [$^\circ$C]}
\nomenclature[V]{$R_{\text{ext}}$}{External thermal resistance (insulation + soil) [K/W]}
\nomenclature[V]{$C_{\text{DHN}}$}{Thermal capacitance of DHN water [kWh/K]}
\nomenclature[V]{$T_{\text{ext}}$}{Effective ground temperature at pipe depth [$^\circ$C]}
\nomenclature[V]{$T_{\text{DHN,min}},T_{\text{DHN,max}}$}{Admissible DHN temperature bounds [$^\circ$C]}

\nomenclature[V]{$p_{e,n}$}{Vector of nodal active power injections [kW]}
\nomenclature[V]{$p_{e,e}$}{Vector of line active power flows [kW]}
\nomenclature[V]{$b_i$}{Line susceptance of edge $i$ [p.u.]}
\nomenclature[V]{$\tilde{F}_e$}{DC power flow mapping matrix}
\nomenclature[V]{$u_{\text{PCC}}$}{Active power exchange at PCC [kW]}
\nomenclature[V]{$d_{\text{PV}}$}{PV active power injection [kW]}
\nomenclature[V]{$d_{\text{Load}}$}{Aggregated electrical load [kW]}

\nomenclature[V]{$c_{\text{el}}$}{Electricity price signal [€/MWh]}
\nomenclature[V]{$c_{\mathrm{gas}}$}{Gas price [€/kg]}
\nomenclature[V]{$\text{LHV}$}{Lower heating value of fuel [MWh/unit]}

\section{Introduction}
\label{sec:intro_related}

Heating and cooling remain one of the major challenges on the path toward deep decarbonization.
In Germany, these sectors accounted for 56 \% of final energy consumption in 2024, yet renewables supplied only 18.2 \% of this demand, far below their approximately 62 \% share in the electricity sector \citep{ArbeitsgemeinschaftEnergiebilanzen2025,UWBA}.
This disparity shows how strongly heat generation still depends on fossil fuels, which poses a critical barrier to national and global climate targets \citep{lund_4th_2014}.

At the same time, thermal networks offer large-scale inherent storage by thermal inertia, whereas power grids lack such buffering capacity.
Exploiting these complementary characteristics is a direct route to system-level flexibility and stronger renewable integration.

Coordinated operation across thermal and electrical domains, however, introduces several challenges.
The volatility of renewable generation, uncertainties in electrical and thermal demand, and the individual dynamics of units such as combined heat and power (CHP) plants, heat pumps (HPs), batteries, and thermal energy storage (TES) complicate the operational decision-making.
Coordinating them means handling assets on different time scales without violating network constraints, as in campus-scale systems that couple CHP, thermal storage, photovoltaic (PV), and waste-heat-driven heat pumps through shared thermal and electrical networks.

Conventional approaches to operating such systems, including rule-based dispatch and static optimization, typically treat assets in isolation or rely on fixed schedules, and therefore lack the ability to anticipate future states, price signals, or renewable availability. As system complexity grows, these strategies leave much of the cost-reduction potential untapped and struggle to enforce coupled network constraints across multiple energy carriers.

These limitations motivate Model Predictive Control (MPC), which optimizes control inputs over a receding horizon while explicitly incorporating system dynamics, network constraints, and forecasts of renewable generation and demand. Economic MPC (EMPC), in particular, directly minimizes operational costs rather than tracking predefined setpoints, an important distinction for multi-energy systems where cost-optimal operation is inherently time-varying. While MILP-based scheduling approaches handle cost-optimal coordination effectively in open-loop settings, they yield fixed schedules that cannot react to forecast errors or disturbances at execution time. This inability to close the loop motivates the receding-horizon EMPC formulation developed here.
This work therefore builds a single EMPC framework that optimizes the thermal and electrical subsystems jointly and still solves fast enough for real-time use. That requires reduced-order yet accurate dynamic models of the thermal and electrical networks, an economic cost function covering fuel and electricity costs alongside renewable utilization, and an optimization structure that stays feasible and economically consistent under time-varying disturbances.

\subsection{Related Work}
Optimal scheduling of multi-energy systems has a long tradition in operations research (OR) and energy economics. Mixed-integer linear programming (MILP) formulations for economic dispatch and unit commitment have been widely applied to coordinate distributed generation, storage, and demand-side resources over day-ahead or intra-day horizons \citep{merkert2020optimal, aguilera2024milp}.
Foundational work in this area includes that of Casisi et al.~\citep{casisi2009optimal}, who developed a MILP model jointly optimizing the layout and operation of a distributed CHP system with microturbines and district heating for urban public buildings. Subsequent studies advanced MILP dispatch with detailed pipeline dynamics \citep{merkert2020optimal}, performance degradation via digital twins \citep{aguilera2024milp}, portfolio optimization comparing merit order and MILP \citep{gonzalez2023portfolio}, and multilevel stochastic frameworks incorporating N-1 security \citep{hu2024multilevel}. Reviews such as \citep{wang2019review} and \citep{sun2024day} synthesize cost- and market-based CHP scheduling, emphasizing thermal inertia and multi-type demand response for wind integration. Tools such as DER-CAM \citep{mashayekh_mixed_2017} and scheduling
frameworks such as those developed by
\citep{schulz_more_2020, schick_role_2022} provide established
benchmarks for cost-optimal energy management. In particular, Schulz et al.~\citep{schulz_more_2020} propose an integrated two-part control for heat pumps coupling optimization scheduling with detailed simulation, while Schick et al.~\citep{schick_role_2022} quantify prosumer self-consumption impacts in near-100\% RES systems, revealing flexibility redundancies and distributional effects from regulatory levies. These offline approaches yield open-loop schedules requiring lower-level execution. In contrast, the EMPC formulation developed here closes the loop via continuous reoptimization with state feedback, providing robustness to forecast errors while matching economic objectives. For central energy
plants with discrete on/off equipment decisions and time-varying
electricity tariffs, Risbeck et al.~\citep{risbeck2017milp} developed a
MILP-based receding-horizon formulation for real-time HVAC
dispatch, demonstrating cost reduction
over heuristic scheduling and over a continuous
relaxation without discrete variables, illustrating the
economic value of retaining binary commitment decisions within
the optimization layer.

In electrical systems, aggregation concepts such as Virtual Power Plants combine distributed generation units, controllable loads, and storage into a coordinated entity \citep{saboori_virtual_2011}. Dynamic Virtual Power Plants extend this idea to provide fast ancillary services and short-term flexibility \citep{marinescu_dynamic_2022, haberle_control_2022}. However, as Abdelkader et al.~\citep{abdelkader_virtual_2024} highlight, these concepts remain almost entirely electrical: thermal assets such as HPs, HVAC systems, and DHNs are rarely included, even though they offer large amounts of slow-timescale flexibility. Co-optimizing thermal and electrical assets under MPC reaches flexibility that purely electrical implementations cannot.

MPC is established practice for energy management in dynamic systems \citep{samad_industry_2020}. In DHNs, it has been applied to multi-source systems with day-ahead price optimization \citep{descamps2019operational}, robust control under uncertainty \citep{farahani2017robust}, mixed-integer unit commitment \citep{hering2021temperature}, network model uncertainty \citep{quaggiotto2021management}, multi-agent scalability \citep{saletti2020development}, building thermal mass as storage \citep{vanhoudt2018active}, and data-driven forecasting using neural networks \citep{verrilli2017model} or physics-informed approaches \citep{de_giuli_physics-informed_2024}.

Recent studies have increasingly targeted integrated thermal--electric systems. Rose et al.~\citep{rose_predictive_2023} coupled DC power flow models with reduced-order thermal networks for real-time MPC in electro-thermal microgrids, demonstrating computational feasibility for operational horizons. Behrunani et al.~\citep{behrunani_distributed_2024} formulated a distributed economic MPC for interconnected energy hubs, co-optimizing thermal and electrical flows via multi-horizon decomposition to maintain long-term planning without excessive computation.

Further contributions include device-level distributed MPC for coordinated CHP, HP, battery, and PV operation \citep{el-afifi_coordinated_2024}, co-simulation frameworks combining high-fidelity network models with low-fidelity MPC \citep{leitner_control_2020}, stochastic MPC for CHP-powered district heating \citep{verrilli_stochastic_2016}, and successive linearization for fast thermal--electric coordination \citep{hoshino_model_2024}.

\subsection{Contribution}
Unified state-space frameworks that couple DHNs and power grids for
joint optimization are still rare. The co-optimization of thermal inertia
and battery storage, despite their complementary time scales, has
received comparatively little attention, as has the systematic
integration of real-time market signals into the MPC cost function.
High-fidelity multi-energy models further impose a heavy
computational burden, motivating reduced-order approximations for
operational use.

\begin{figure}
    \centering
    \includegraphics[width=\linewidth]{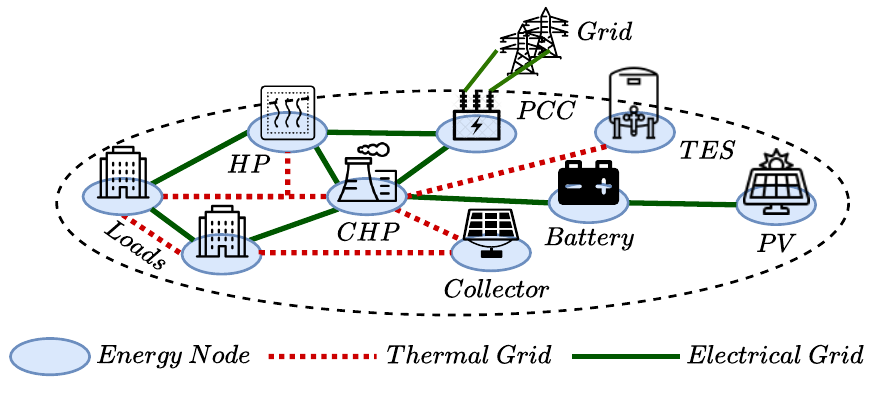}
    \caption{Schematic representation of the multi-energy system integrating thermal and electrical flows, storage systems, and renewable energy sources for optimized operation.}
    \label{fig:overview}
\end{figure}

This work addresses these gaps with a unified EMPC framework for
integrated multi-energy grids that jointly operates CHPs, large-scale
HPs, TES, and batteries within a single mixed-integer state-space
formulation. The cost function is convex and captures fuel-electricity
trade-offs, grid exchange costs, and renewable curtailment, while
accommodating time-varying market signals. Reduced-order network models keep the resulting program small enough
to solve online:
resistive--capacitive DHN representations \citep{felczak_dynamic_2019},
linearized storage dynamics \citep{de_lorenzi_predictive_2022}, and DC
power flow \citep{purchala_usefulness_2005,rose_predictive_2023}. The
overall system topology is illustrated in Figure~\ref{fig:overview}.

The main contributions of this work are as follows:
\begin{itemize}
\item A unified, affine state-space model that couples an RC-based
district heating network (DHN) with a DC power flow electrical grid, five
generation/storage components, and time-varying disturbances in a
single mixed-integer linear program with a convex economic stage cost.

\item A cyclic-terminal EMPC with scaled soft terminal constraints
that anchors receding-horizon operation to an offline periodic
economic orbit, exploiting daily price and demand periodicity without
hard terminal sets.

\item A practical mixed-integer EMPC implementation for a
campus-scale multi-energy system, integrating CHP units with Mixed Logical Dynamical (MLD)
startup logic, a large-scale heat pump, thermal energy storage, a
battery, and network constraints.

\item

A closed-loop sensitivity analysis characterizing two
structural design properties: (i) a horizon sweep run with and without
the terminal condition, showing that the prediction horizon and the cyclic
terminal condition act as substitutes: without anchoring, the cost gap to the
window-matched periodic reference closes only slowly and monotonically with
lookahead, whereas with a sufficiently large terminal weight the gap is small
and near-flat across the entire swept range, down to the shortest horizon
tested, with the residual spread attributable to integer-switching noise, a
result cross-validated on a second, independent price week,
and (ii) a terminal weight sweep
identifying a low-weight regime in which the soft terminal
constraint is inactive, an activation threshold beyond which it
engages, and a wide multi-decade cost plateau thereafter within
which the baseline choice sits comfortably.

\item
An empirical characterization of a receding-horizon
drift of the thermal energy storage, a cost-neutral storage
state that the offline periodic reference keeps inert but the
online EMPC drives far across its range, together with an
interpretation consistent with a failure of strict dissipativity
in that coordinate, under which average-performance guarantees
hold for the hard-constrained formulation of
\mbox{
\citep{risbeck_economic_2020} }\hskip0pt
while trajectory convergence to the
reference is not implied; the implemented soft-penalty scheme trades
that formal bound for feasibility robustness (Section~\ref{sec:methodology}).
\end{itemize}

The paper is organized as follows: Section~\ref{sec:background}
introduces time-varying EMPC theory; Section~\ref{sec:methodology}
presents the component models, the unified formulation, the EMPC
design, and the experimental setup; Section~\ref{sec:results}
discusses results; and Section~\ref{sec:conclusion} concludes.

\section{Economic MPC with time-varying cost function}
\label{sec:background}


Both tracking MPC and EMPC minimize a finite-horizon cost of the form
\begin{equation}
    J_N(\mathbf{x}, \mathbf{u}, t)
    = \sum_{k=0}^{N-1} \ell\big(\mathbf{x}(k), \mathbf{u}(k), t+k\big)
    + J_f\big(\mathbf{x}(N), t+N\big);
    \label{eq:finite_horizon_cost}
\end{equation}
the distinction lies in the choice of the stage cost $\ell$ and
terminal cost $J_f$. Tracking MPC penalizes the deviation from a
predefined setpoint or reference trajectory, whereas EMPC employs the
economic objective itself, such as energy cost or fuel consumption,
as the stage cost \citep[p.153]{rawlings_model_2017}. This section
recalls the elements of the time-varying EMPC framework of
\citep{risbeck_economic_2020} that form the basis of the cyclic-terminal
formulation used in this work.

The plant is described by linear time-invariant continuous dynamics
\begin{equation}
\dot{\mathbf{x}}(t) = A\, \mathbf{x}(t) + B\, \mathbf{u}(t)
+ E\, \mathbf{d}(t) + w,
\label{eq:state_space}
\end{equation}
where $A$, $B$, and $E$ map the state $\mathbf{x}$, input
$\mathbf{u}$, and disturbance $\mathbf{d}$, and $w \in
\mathbb{R}^{n_x}$ is a constant affine term. The affine term
$w$ is a \emph{known}, constant vector (not a disturbance) collecting
the fixed physical offsets of the component models, such as the TES
lower temperature $T_L$ and the DHN ground temperature
$T_{\mathrm{ext}}$. It acts only on the strictly stable thermal states
and is retained explicitly so that each offset stays traceable to its
originating component. For controller design,
\eqref{eq:state_space} is discretized by zero-order hold at sampling
period $T_s$ (Section~\ref{sec:integrated}), yielding
\begin{equation}
\mathbf{x}^{+} = A_d\,\mathbf{x} + B_d\,\mathbf{u} + E_d\,\mathbf{d}
+ w_d,
\label{eq:dt_dynamics_bg}
\end{equation}
where $\mathbf{x}^{+}$ denotes the successor state. After
discretization, $t \in \mathbb{I}_{\ge 0}$ denotes the discrete
time index (corresponding to sampling instant $t\,T_s$), and $k$
the prediction-step offset within the horizon. The stage cost
$\ell(\mathbf{x},\mathbf{u},t)$ is time-varying even though the
dynamics are not, because electricity prices and demand profiles
vary with time.

In the energy systems considered here, prices and demand exhibit a
strong, but not exact, daily pattern. We therefore do not assume the
disturbances themselves to be periodic. Instead, a nominal
disturbance estimate $\{\hat{d}_k\}_{k=0}^{N_p-1}$ over one period
of length $N_p$ is used to construct, offline, an optimal
periodic orbit $\{x_k^{\mathrm{ref}},
u_k^{\mathrm{ref}}\}_{k=0}^{N_p-1}$: the state--input sequence
minimizing the economic cost over one period subject to the
dynamics, the constraints, and the periodic boundary condition
$x_{N_p} = x_0$ (stated formally in
Section~\ref{sec:empc_formulation},
Eq.~\eqref{eq:offline_ref}). Extended $N_p$-periodically, this orbit
is a feasible trajectory of the nominal system in the sense of
\citep[Assumption~1]{risbeck_economic_2020}.
It is computed once offline and held fixed throughout the
closed-loop run; it is never re-solved or updated online and therefore
introduces no jumps. Being a feasible trajectory, it satisfies the
system dynamics at every step, and the online controller only slides
its terminal anchor along this fixed reference as time advances.
We emphasize that the
orbit is not an exogenous tracking target: it is generated from the
same economic stage cost and dynamics as the online controller, and
it enters the online problem only through the terminal ingredients
introduced below. The intra-horizon behavior remains governed purely
by the economic stage cost.

The role of the orbit is justified by Theorem~1 of
\citep{risbeck_economic_2020}: if the reference is a feasible
trajectory and the terminal ingredients satisfy a recursive
feasibility condition
(Assumption~3 of \citep{risbeck_economic_2020}), the
asymptotic average closed-loop cost of
the receding-horizon controller does not exceed the average cost of
the reference.
For the exact terminal equality $\mathcal{X}_f(t) =
\{x^{\mathrm{ref}}(t)\}$ with $V_f\equiv 0$, this condition is met by
$u=u^{\mathrm{ref}}(t)$, which keeps the successor state on the
reference and renders the terminal-cost decrease an equality; it thus
follows from reference feasibility (Assumption~1) alone, requiring no
separately constructed terminal cost or invariant set.
Two aspects of this result matter for our setting.

First, the bound requires only that the reference be a
feasible trajectory; it presumes neither periodicity nor any
optimality property of the orbit. The guarantee therefore survives
the use of a nominal periodic orbit even when realized disturbances
deviate from the estimate, provided the orbit remains feasible for
the realized system. Second, it accommodates discrete-valued
actuators: the input sets are required only to be compact, not
convex or with nonempty interior
\citep[Remark~2]{risbeck_economic_2020}, so the binary CHP
commitment variables of Section~\ref{sec:models} are covered. The
result is verified for the exact terminal equality
$\mathcal{X}_f(t) = \{x^{\mathrm{ref}}(t)\}$; our implementation
softens this constraint, as discussed below, and we assess the
resulting orbit alignment empirically in
Section~\ref{sec:terminal_sweep}.

A complementary property explaining the practical effectiveness of
finite horizons is the \emph{turnpike property}: optimal
trajectories spend the majority of the horizon in the vicinity of
the optimal orbit, deviating only near the initial and terminal
steps \citep{grune_economic_2013}, so that the closed-loop average
cost approaches the orbit cost as the horizon $N$ grows.
The same turnpike mechanism also underlies performance and
stability guarantees for periodic economic MPC \emph{without} terminal
conditions, such as the linearly discounted scheme of
\citep{schwenkel_linearly_2024}; we instead follow the
terminal-condition route, anchoring the finite horizon to the
precomputed periodic orbit.
We use
this property empirically in Section~\ref{sec:horizon_sweep}, where
the per-state turnpike fraction gives a direct measure of how well
the online controller tracks the periodic orbit across horizon
lengths.

Building on these elements, the finite-horizon EMPC problem solved
at time $t$ is
\begin{align}
\min_{\substack{\{x,\,u\},\\ s_N\ge 0}} \quad &
\sum_{k=0}^{N-1} \ell \big(x(k),u(k),t+k\big)
+ \rho_N^\top s_N, \label{eq:empc_obj_final}\\[2mm]
\text{s.t.}\quad
& x(0) = x(t), \quad x^{+} = A_d x + B_d u + E_d d + w_d,
\label{eq:empc_dynamics}\\
& x \in \mathcal{X}, \quad u \in \mathcal{U}, \nonumber\\
& |C_T x_N - r_N| \le s_N \quad \text{(componentwise)},
\label{eq:tube_final}
\end{align}
with terminal reference $r_N = C_T\,x_{t+N}^{\mathrm{ref}}$, where
$C_T \in \mathbb{R}^{n_s \times n_x}$ selects the $n_s$ terminal
states of interest, $s_N \in \mathbb{R}^{n_s}_{\ge 0}$ are slack
variables, and $\rho_N \in \mathbb{R}^{n_s}_{>0}$ penalizes terminal
deviations from the periodic orbit. Relative to the exact terminal
equality of \citep{risbeck_economic_2020}, the slack formulation
trades the formal performance bound for robustness: when realized
disturbances render the nominal orbit unreachable within the
horizon, the soft tube preserves feasibility of the online problem
while still anchoring the terminal state to the orbit.

Absent binary variables, linear dynamics and a convex stage cost
render \eqref{eq:empc_obj_final}--\eqref{eq:tube_final} a convex
program. With the binary commitment variables of the hybrid CHP
formulation (Section~\ref{sec:models}), the problem becomes a
mixed-integer linear program whose continuous relaxation retains
convexity. The controller implemented in
Section~\ref{sec:empc_formulation} instantiates this structure with
a linear slack penalty and individually scaled terminal constraints.

\begin{figure}[h!]
    \centering
    \includegraphics[width=1\linewidth]{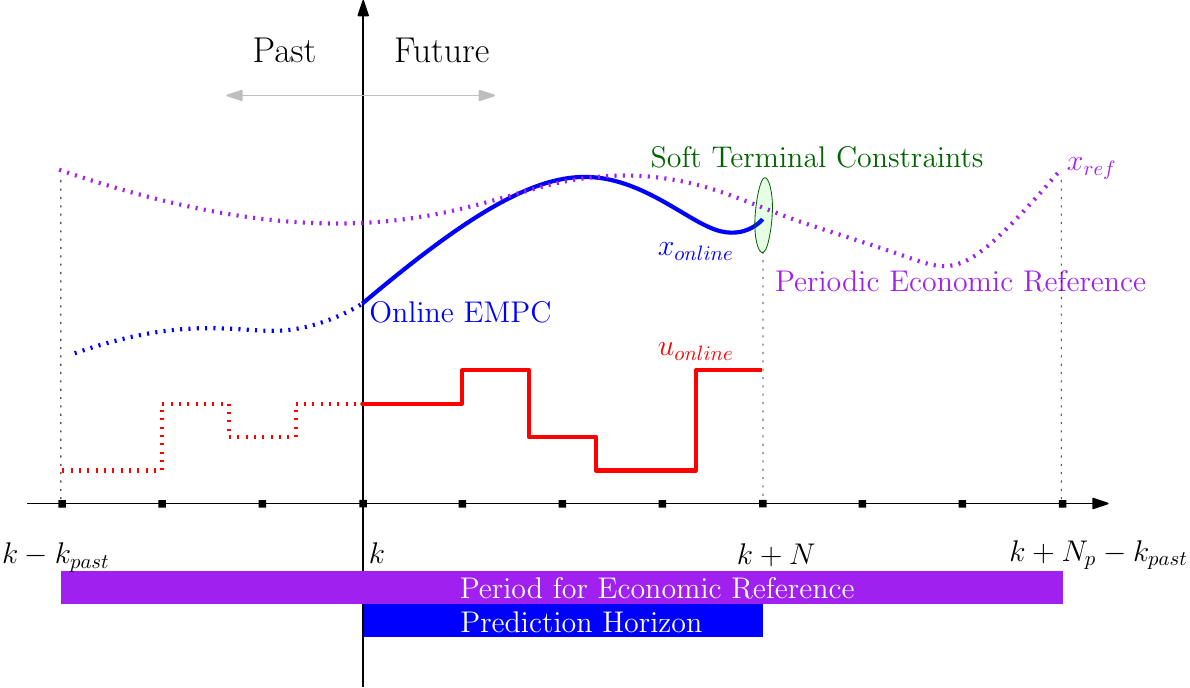}
    \caption{Conceptual illustration of the time-varying EMPC scheme
    with a periodic economic reference. The EMPC optimizes the
    future trajectories \(x_{\text{online}}\) and
    \(u_{\text{online}}\) within the prediction horizon \(N\), while
    the terminal tube constrains the terminal state around the
    periodic reference \(x_{\mathrm{ref}}(t)\).}
    \label{fig:empc_ifac}
\end{figure}

Figure~\ref{fig:empc_ifac} illustrates the scheme: at each sampling
instant, the controller optimizes over the prediction horizon
subject to the soft terminal tube, applies the first control action,
and shifts the horizon forward, producing a closed-loop trajectory
that converges toward the periodic operating orbit.

\section{Methodology}
\label{sec:methodology}
The proposed EMPC framework places the thermal and electrical subsystems in one state-space representation, so both are optimized against the same cost. This section summarizes the key component models and the overall formulation.

\subsection{Mathematical Formulation of Components}
\label{sec:models}

After describing the investigated multi-energy system, we present the
mathematical models for our EMPC framework. The goal is to derive one
state-space representation comprising all components and grids in the form
of equation~\eqref{eq:state_space}, where $\mathbf{x}(t) \in \mathbb{R}^{n_x}$
is the state vector (e.g., stored energy in thermal storage or
battery state-of-charge); $\mathbf{u}(t) \in \mathbb{R}^{n_u}$ is the
control input vector (e.g., power setpoints, heat input); $\mathbf{d}(t)
\in \mathbb{R}^{n_d}$ represents exogenous disturbances (e.g., ambient
temperature, solar irradiation). Matrices $A \in \mathbb{R}^{n_x \times n_x}$,
$B \in \mathbb{R}^{n_x \times n_u}$, and $E \in \mathbb{R}^{n_x \times n_d}$
describe the component's intrinsic dynamics, control input effects, and
disturbance impacts, respectively; the constant affine term
$w \in \mathbb{R}^{n_x}$ collects constant contributions from individual
component models.

The multi-energy system considered in this work comprises five major components, a stratified TES, a battery, a large-scale HP, a CHP plant, and a DHN, interconnected through thermal and electrical distribution grids as depicted in Fig.~\ref{fig:overview}. The following subsections present the dynamic model of each component, followed by the disturbance model, the DHN and electrical grid formulations, and the integrated state-space assembly.

Throughout this section, each physical component is derived in
continuous time using the notation $\dot{x}(t)$. The combined
model is discretized once in Section~\ref{sec:integrated}
(Eq.~\eqref{eq:global_dt}), after which the discrete successor
notation $x^+$ is used consistently. The CHP is the only
component formulated directly in discrete time, as its hybrid
on/off logic is intrinsically discrete.

\subsubsection{Thermal Energy Storage}
The TES is modeled via a linear thermocline-based state-space framework adapted from \cite{de_lorenzi_predictive_2022}, using the stored thermal energy $Q_{\text{TES}}(t)$ as the state variable instead of the thermocline height. After combining losses from the high- and low-temperature zones through the tank surface area $S_{\text{TES}} = \pi D_{\text{TES}} H_{\text{TES}}$, where $D_{\text{TES}}$ and $H_{\text{TES}}$ are the tank diameter and height, the TES dynamics take the form
\begin{equation}
\dot{x}_{\text{TES}}(t)
=
A_{\text{TES}}x_{\text{TES}}(t)
+
B_{\text{TES}}u_{\text{TES}}(t)
+
E_{\text{TES}} T_{\text{A}}(t)
+
w_{\text{TES}},
\end{equation}
with
\begin{equation}
\begin{aligned}
A_{\text{TES}} &= -\frac{U_{\text{TES}}S_{\text{TES}}(T_H - T_L)}{Q_{\text{TES,N}}}, \\[3pt]
B_{\text{TES}} &= -1, \quad
E_{\text{TES}} = U_{\text{TES}}S_{\text{TES}}, \\
w_{\text{TES}} &= -U_{\text{TES}}S_{\text{TES}}\,T_L.
\end{aligned}
\label{eq:TES_params}
\end{equation}
Here, $x_{\text{TES}}(t) = Q_{\text{TES}}(t)$ is the state,
$u_{\text{TES}}(t) = \dot{Q}_{\text{HS}}(t)$ the charge/discharge power
(negative for charging), $T_A(t)$ the ambient temperature disturbance,
$U_{\text{TES}}$ the overall heat-transfer coefficient of the tank wall,
and $Q_{\text{TES,N}}$ the nominal storage capacity. The parameters $T_H$ and $T_L$ denote the high- and low-temperature levels of the stratified storage, respectively. The lower storage temperature $T_L$ is absorbed into the constant affine term $w_{\text{TES}}$. The discretized form used in the MPC follows from Equation~\eqref{eq:global_dt}. Denoting the maximum charge/discharge power by $\dot{Q}_{\max}$,
power and energy are bounded by
\begin{equation}
-\dot{Q}_{\max} \;\le\; \dot{Q}_{\text{HS}}(t) \;\le\; \dot{Q}_{\max},
\label{eq:tes_power_limits}
\end{equation}
\begin{equation}
0 \;\le\; Q_{\text{TES}}(t) \;\le\; Q_{\text{TES,N}}.
\label{eq:tes_energy_limits}
\end{equation}

\subsubsection{Battery Storage Model}
The battery model neglects cycling losses, capacity fade, and thermal aging effects. With state $x_{\text{batt}}(t) = \text{SOC}$ and inputs $u_{\text{batt}}(t) = [P_{\text{ch}}(t),\; P_{\text{disch}}(t)]^\top$, the dynamics are
\begin{equation}
    \dot{x}_{\text{batt}}(t) =
    \underbrace{\begin{bmatrix}
    \dfrac{\eta_{\text{ch}}}{C_{\text{batt}}} &
    -\dfrac{1}{\eta_{\text{disch}}\,C_{\text{batt}}}
    \end{bmatrix}}_{B_{\text{batt}}}
    u_{\text{batt}}(t),
    \label{eq:batt_dynamics}
\end{equation}
where $\eta_{\text{ch}}$ and $\eta_{\text{disch}}$ are the charging and
discharging efficiencies, and $C_{\text{batt}}$ is the battery energy capacity.
A binary variable $\gamma_{\text{batt}}(t) \in \{0,1\}$ enforces mutually
exclusive charging and discharging, with $P_{\text{batt,max}}$ the maximum
battery power:
\begin{equation}
\begin{aligned}
0 &\le P_{\text{ch}}(t) \le P_{\text{batt,max}}\,\gamma_{\text{batt}}(t), \\[2pt]
0 &\le P_{\text{disch}}(t) \le P_{\text{batt,max}}\,[1 - \gamma_{\text{batt}}(t)].
\end{aligned}
\label{eq:batt_constraints}
\end{equation}

The SOC is bounded by $\mathrm{SOC}_{\min} \le x_{\text{batt}}(t) \le \mathrm{SOC}_{\max}$,
where $\mathrm{SOC}_{\min}$ and $\mathrm{SOC}_{\max}$ are the lower and upper
admissible charge limits.
The net active power injection at the battery's electrical node is
\begin{equation}
    y_{\text{batt}}(t) =
    \underbrace{\begin{bmatrix}
    -1 & 1
    \end{bmatrix}}_{D_{\text{batt}}}
    u_{\text{batt}}(t)
    = P_{\text{disch}}(t) - P_{\text{ch}}(t),
    \label{eq:batt_output}
\end{equation}
which enters the nodal power injection vector of the electrical layer. Unlike a full MLD formulation~\citep{bemporad_control_1999}, the binary variable only affects the admissible input set, keeping the model efficient for mixed-integer MPC.

\subsubsection{Large Scale Heat Pump Model}
Heat pumps convert electrical power and low-temperature heat sources
into elevated-temperature thermal energy. To capture the dynamic behavior of large-scale HPs, we identified a first-order model from step-response measurements of a two-stage screw/piston HP with a nominal thermal output of approximately 1.5 MW \citep{agfw2024stromnetz}. A Pad\'{e}-approximated model was also fitted and provides an improved transient match, but the first-order model is used here for its favorable balance of accuracy and computational efficiency.
The identified parameters are $K=1$ and $\tau=247.46$ s. The continuous-time state-space representation is
\begin{equation}
\begin{aligned}
\dot{x}_{\text{HP}}(t) &= A_{\text{HP}}\,x_{\text{HP}}(t) + B_{\text{HP}}\,u_{\text{HP}}(t), \\
y_{\text{HP}}(t) &= C_{\text{HP}}\,x_{\text{HP}}(t),
\end{aligned}
\end{equation}
where $x_{\text{HP}}(t)$ is the internal state, $u_{\text{HP}}(t) \in [0,1]$ the normalized electrical input, and $y_{\text{HP}}(t)$ the normalized output power. Denoting the rated electrical power of the heat pump by $P_{\text{HP,el}}^{\max}$,
the electrical power consumption is
\begin{equation}
    P_{\text{HP,el}} = P_{\text{HP,el}}^{\max}\, C_{\text{HP}}\, x_{\text{HP}}, \qquad
    0 \;\le\; P_{\text{HP,el}}\;\le\; P_{\text{HP,el}}^{\max},
\end{equation}
and the thermal output is obtained through a constant COP representative
of the operating range:
\begin{equation}
    P_{\text{HP,th}} = \mathrm{COP} \cdot P_{\text{HP,el}}, \qquad
    P_{\text{HP,th}} \ge 0.
\end{equation}
The constant COP of $3.252$ is determined in two steps. First, following
\cite{jesper_large-scale_2021}, the reference COP depends primarily on the
temperature lift $\Delta T_{\text{lift}} = T_{\text{sink}} - T_{\text{source}}$
between the heat sink and source. At the considered operating point
($\Delta T_{\text{lift}} = 55$~K, $T_{\text{source}} = 45$~°C), this yields
$\mathrm{COP}_{\text{ref}} = 3.47$. Second, for part-load operation the
effective COP follows a quadratic correction with respect to the load ratio,
reaching its maximum at approximately 80\% of rated capacity. Evaluating
this part-load characteristic once at the nominal operating point gives the
constant $3.252$ used throughout this work. To maintain
linearity within the MPC framework, temperature-dependent COP variations
are not modeled explicitly; temperature-dependent and
Pad\'{e}-model extensions are left for future work.

\subsubsection{Mixed Logical Dynamical CHP Model}

The CHP is formulated in the discrete-time Mixed Logical Dynamical
(MLD) framework~\citep{bemporad_control_1999}, based
on~\cite{weber_realistic_2018}, to represent its hybrid on/off
dynamics within mixed-integer MPC. Unlike the continuous-time
components discretized via ZOH in Section~\ref{sec:integrated},
the CHP is defined directly in discrete time at sampling period
$T_s$.

The CHP state vector collects a continuous and a binary component:
\[
x_{\mathrm{CHP}} =
\begin{bmatrix} x_c & x_\ell \end{bmatrix}^\top
=
\begin{bmatrix} P_{\mathrm{set}} & \delta^{\mathrm{on}} \end{bmatrix}^\top,
\qquad
x_c\in\mathbb{R},\;
x_\ell\in\{0,1\},
\]
where $P_{\mathrm{set}}$ is the electrical set power and
$\delta^{\mathrm{on}}$ the on/off status. The input vector
combines continuous and logical control actions:
\[
u_{\mathrm{CHP}} =
\begin{bmatrix} u_c & u_{\mathrm{on}} & u_{\mathrm{st}} \end{bmatrix}^\top
=
\begin{bmatrix} \Delta P & \delta^{\mathrm{on,cmd}} & \delta^{\mathrm{start}} \end{bmatrix}^\top,
\]
with $u_c\in\mathbb{R}$ and $u_{\mathrm{on}},u_{\mathrm{st}}\in\{0,1\}$,
where $\Delta P$ is the power gradient command,
$\delta^{\mathrm{on,cmd}}$ the binary on-command, and
$\delta^{\mathrm{start}}$ the startup indicator.

The one-step dynamics follow the canonical MLD form:
\begin{equation}
x^+_{\mathrm{CHP}} =
\underbrace{\begin{bmatrix}1 & 0\\[2pt] 0 & 0\end{bmatrix}}_{A_{\mathrm{CHP}}}
x_{\mathrm{CHP}}
+
\underbrace{\begin{bmatrix}1 & 0 & 0\\[2pt] 0 & 1 & 0
\end{bmatrix}}_{B_{\mathrm{CHP}}}
u_{\mathrm{CHP}},
\label{eq:chp_mld_dyn}
\end{equation}
where $x^+_{\mathrm{CHP}}$ denotes the successor state at the
next sampling instant; $x_c$ evolves by the power gradient $u_c$,
and $x_\ell$ is updated directly by the on-command $u_{\mathrm{on}}$.

The logical startup behavior is enforced by:
\[
u_{\mathrm{st}} \ge u_{\mathrm{on}} - x_\ell, \quad
u_{\mathrm{st}} \le u_{\mathrm{on}}, \quad
u_{\mathrm{st}} \le 1 - x_\ell,
\]
so that $u_{\mathrm{st}}=1$ only on an off-to-on transition.
Operational constraints enforce feasible power levels and ramp
rates:
\[
P_{\mathrm{CHP}}^{\min}\,x^+_\ell \;\le\; x^+_c
\;\le\; P_{\mathrm{CHP}}^{\max}\,x^+_\ell,
\]
\[
\Delta P_{\min} \;\le\; u_c \;\le\;
\Delta P_{\max} +
\bigl(\Delta P_{\mathrm{start}} - \Delta P_{\max}\bigr)
u_{\mathrm{st}},
\]
with $P_{\mathrm{CHP}}^{\min}$, $P_{\mathrm{CHP}}^{\max}$ the
power limits, $\Delta P_{\min}$, $\Delta P_{\max}$ the ramp-rate
bounds under normal operation, and $\Delta P_{\mathrm{start}}$
the reduced ramp cap applied during a startup transition.

The electrical and thermal outputs are:
\begin{equation}
y_{\mathrm{CHP}} =
\underbrace{\begin{bmatrix}
1 & 0\\[2pt]
\kappa & 0
\end{bmatrix}}_{c_{\mathrm{CHP}}}
x_{\mathrm{CHP}}
+
\underbrace{\begin{bmatrix}
1+\alpha & 0 & P_{\mathrm{start}}\\[2pt]
\kappa(1+\alpha) & 0 & \kappa P_{\mathrm{start}}
\end{bmatrix}}_{D_{\mathrm{CHP}}}
u_{\mathrm{CHP}},
\label{eq:chp_mld_output}
\end{equation}
where $y_{\mathrm{CHP}} = [P_{\mathrm{av}},\;
Q_{\mathrm{chp}}]^\top$, $\kappa = \eta_{\mathrm{th}} /
\eta_{\mathrm{el}}$ links thermal and electrical efficiencies,
$\alpha = (\Delta P_{\max} - \Delta P_{\min}) /
[2(\Delta P_{\max} + \Delta P_{\min})]$ captures the inertia-induced
lag, and $P_{\mathrm{start}}$ is the startup power deficit.
Nonnegativity is enforced by $P_{\mathrm{av}} \ge 0$.

\subsubsection{Disturbances}
\label{sec:disturbances}

All exogenous time-varying signals are treated as disturbances, incorporated either in the system dynamics or the cost function.

To integrate the solar collector into the MPC, the thermal power output $\dot{Q}_{\text{coll}}(t)$ is expressed using the Hottel--Whillier equation~\cite{osti_5057828} as
\begin{equation}
    \dot{Q}_{\text{coll}}(t) = A_{\text{coll}}
    \left(\eta_0 G_{\text{coll}}(t) - a_1 \Delta T(t)\right),
    \label{eq:hottel_willier}
\end{equation}
where $G_{\text{coll}}(t)$ is the incident solar irradiance,
$\Delta T(t) = T_{\text{coll,avg}}(t) - T_{\text{A}}(t)$ the effective
temperature difference between the collector and the ambient air,
$\eta_0$ the optical efficiency, and $a_1$ the first-order heat-loss
coefficient of the collector.
Expanding \eqref{eq:hottel_willier} yields
\begin{equation}
\dot{Q}_{\text{coll}}(t)
= A_{\text{coll}}\eta_0 G_{\text{coll}}(t)
+ A_{\text{coll}} a_1 T_{\text{A}}(t)
- A_{\text{coll}} a_1 T_{\text{coll,avg}}(t),
\label{eq:qcoll_expanded}
\end{equation}
which can be written in the compact affine form
\begin{equation}
\begin{aligned}
\dot{Q}_{\text{coll}}(t)
&= E_{\text{coll}}\, d_{\text{coll}}(t) + w_{\text{coll}}, \\[4pt]
E_{\text{coll}}
&=
\begin{bmatrix}
A_{\text{coll}}\eta_0 & A_{\text{coll}} a_1
\end{bmatrix}, \quad
d_{\text{coll}}(t) =
\begin{bmatrix} G_{\text{coll}}(t) & T_{\text{A}}(t) \end{bmatrix}^\top, \\
w_{\text{coll}} &= -A_{\text{coll}} a_1 T_{\text{coll,avg}}.
\end{aligned}
\label{eq:qcoll_affine}
\end{equation}

Thus, only $G_{\text{coll}}(t)$ and $T_{\text{A}}(t)$ enter as measurable external disturbances,
while the effect of $T_{\text{coll,avg}}$, the mean collector temperature
evaluated at the nominal operating point is absorbed into the constant
affine term $w_{\text{coll}}$.
The resulting heat flow $\dot{Q}_{\text{coll}}(t)$ is injected into the DHN as an exogenous input,
allowing the controller to anticipate variations in solar availability and ambient temperature through forecast data.

The electrical power generated by the PV system is modeled as an external disturbance \(d_{\text{PV}}(t)\).
No explicit dynamic model is assumed, and perfect foresight of forecasted PV profiles is assumed throughout this study, consistent with the perfect-foresight setting of Section~\ref{sec:results}.
The net PV injection $p_{\text{PV,net}}(t) = d_{\text{PV}}(t) - p_{\text{curtail}}(t)$ is constrained
by grid capacity $p_{\text{PV,net}}(t) \leq p_{\text{PV}}^{\max}$, where curtailment $p_{\text{curtail}}(t) \geq 0$
is an optimization variable with economic penalties applied during positive-price periods in the cost function.

Thermal and electrical demands of the buildings connected to the power and thermal grid are modeled as time-varying disturbance signals,
\begin{equation}
d_{\text{load}}(t) = \begin{bmatrix} d_{\text{load,th}}(t) & d_{\text{load,el}}(t) \end{bmatrix}^\top,
\label{eq:load_disturbance}
\end{equation}
where \(d_{\text{load,el}}(t)=p_{\text{load}}\) and \(d_{\text{load,th}}(t)=\dot{Q}_{\text{load}}\) denote the forecasted thermal and electrical load demands, respectively.
This formulation allows both domains to be represented compactly while maintaining clear reference to each component for coupling between the electrical and thermal layers.
The ambient temperature \(T_{\text{A}}(t)\) influences both the collector efficiency and the thermal storage heat losses, and thus directly affects the thermal subsystem dynamics.

For compact notation, the overall disturbance vector is separated into
\emph{physical} and \emph{economic} components expressed with
\begin{equation}
\footnotesize
d_{\mathrm{phys}}(t) =
\begin{bmatrix} T_{\!A}(t) & G_{\mathrm{coll}}(t) & \dot{Q}_{\mathrm{load}}(t) & P_{\mathrm{load}}(t) & T_{{\mathrm{ext}}}(t) & d_{\mathrm{PV}}(t) \end{bmatrix}^\top
\label{eq:disturbance_vectors}
\end{equation}
\normalsize
and $d_{\mathrm{econ}}(t) = c_{\mathrm{el}}(t)$.

\subsubsection{District Heating Network Model}
\begin{figure}[ht]
    \centering
    \includegraphics[width=0.6\linewidth]{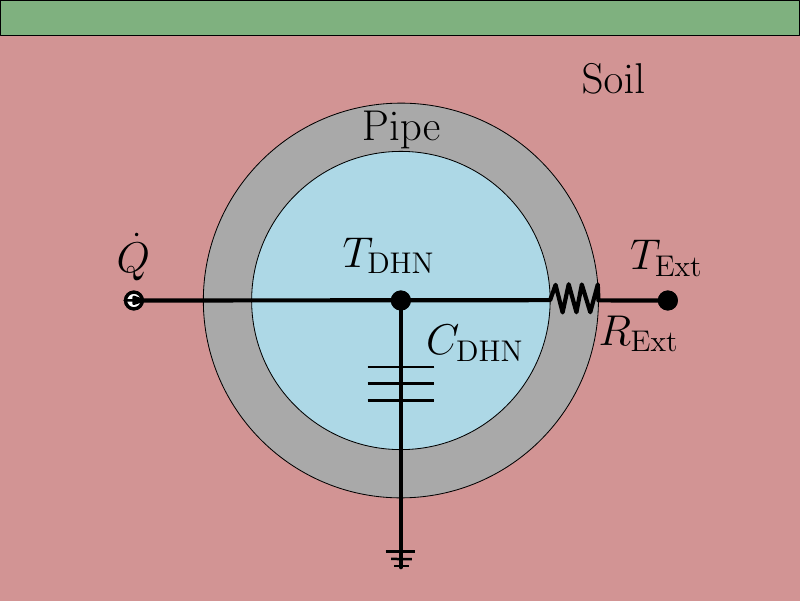}
    \caption{Resistive-capacitive equivalent circuit model for a DHN pipe system.}
    \label{fig:rc_dhn}
\end{figure}

For the DHN, we employ a resistive--capacitive (RC) model adapted from \cite{matthiss_thermal_2023,felczak_dynamic_2019} as shown in Fig. \ref{fig:rc_dhn}. This reduced-order representation neglects hydraulic pressure dynamics and detailed pipe-level heat losses (in contrast to detailed models such as \cite{xu_integrated_2023}), focusing instead on the dominant thermal storage effect of the network water mass. The model comprises an external temperature node $T_{\text{ext}}(t)$ representing the effective ground temperature at pipe burial depth, connected through the equivalent insulation resistance $R_{\text{ext}}$, the effective thermal capacitance of the network water mass $C_{\text{DHN}}$, and an internal water temperature node $T_{\text{DHN}}(t) = (T_{\text{supply}}(t) + T_{\text{return}}(t))/2$.

The net thermal power balance of the grid is expressed as
\begin{align}
\sum_i \dot{Q}_{\text{prod},i}(t) - \sum_j \dot{Q}_{\text{consum},j}(t)
&=
P_{\text{HP,th}}(t)
+ \dot{Q}_{\text{HS}}(t)
+ Q_{\text{chp}}(t) \nonumber \\[3pt]
&\quad
+ \dot{Q}_{\text{coll}}(t)
- \dot{Q}_{\text{load}}(t),
\end{align}
where $P_{\text{HP,th}}(t)$ is the heat pump thermal output,
$\dot{Q}_{\text{HS}}(t)$ the TES charge/discharge power,
$Q_{\text{chp}}(t)$ the CHP thermal output,
$\dot{Q}_{\text{coll}}(t)$ the solar collector heat injection,
and $\dot{Q}_{\text{load}}(t)$ the aggregated consumer heat demand.

Using the RC analogy, the first-order differential equation
governing the DHN temperature dynamics becomes
\begin{equation}
\begin{split}
\dot{T}_{\text{DHN}}(t)
&= \underbrace{-\frac{1}{R_{\text{ext}}C_{\text{DHN}}}}_{A_{\text{DHN}}}
\,T_{\text{DHN}}(t) \\[3pt]
&\quad +
\underbrace{\begin{bmatrix}
\dfrac{1}{C_{\text{DHN}}} &
\dfrac{1}{C_{\text{DHN}}} &
\dfrac{1}{C_{\text{DHN}}}
\end{bmatrix}}_{B_{\text{DHN}}}
\begin{bmatrix}
P_{\text{HP,th}}(t) \\[2pt]
\dot{Q}_{\text{HS}}(t) \\[2pt]
Q_{\text{chp}}(t)
\end{bmatrix} \\[3pt]
&\quad +
\begin{bmatrix}
\dfrac{1}{C_{\text{DHN}}} &
-\dfrac{1}{C_{\text{DHN}}} &
\dfrac{1}{R_{\text{ext}}C_{\text{DHN}}}
\end{bmatrix}
\begin{bmatrix}
\dot{Q}_{\text{coll}}(t) \\[2pt]
\dot{Q}_{\text{load}}(t) \\[2pt]
T_{\text{ext}}(t)
\end{bmatrix}.
\end{split}
\label{eq:dhn_raw}
\end{equation}

Using the affine representations of the solar collector and TES
losses derived in the previous sections, $\dot{Q}_{\text{coll}}(t)$
and $\dot{Q}_{\text{HS}}(t)$ can be expressed as linear functions
of the disturbances $G_{\text{coll}}(t)$ and $T_{\text{A}}(t)$
plus constant offsets. Substituting into \eqref{eq:dhn_raw} and
grouping terms yields the compact affine state-space form
\begin{equation}
\dot{x}_{\text{DHN}}(t)
= A_{\text{DHN}}x_{\text{DHN}}(t)
+ B_{\text{DHN}}u_{\text{DHN}}(t)
+ E_{\text{DHN}}d_{\text{DHN}}(t)
+ w_{\text{DHN}},
\end{equation}
with
\begin{align*}
x_{\text{DHN}}(t) &= T_{\text{DHN}}(t), \\
u_{\text{DHN}}(t) &=
\begin{bmatrix} P_{\text{HP,th}}(t) & \dot{Q}_{\text{HS}}(t) & Q_{\text{chp}}(t) \end{bmatrix}^\top, \\
d_{\text{DHN}}(t) &=
\begin{bmatrix} G_{\text{coll}}(t) & T_{\text{A}}(t) & \dot{Q}_{\text{load}}(t) & T_{\text{ext}}(t) \end{bmatrix}^\top.
\end{align*}
The DHN disturbance matrix and affine term then become
\begin{equation}
\begin{aligned}
E_{\text{DHN}} &=
\begin{bmatrix}
\dfrac{A_{\text{coll}}\eta_0}{C_{\text{DHN}}} &
\dfrac{A_{\text{coll}} a_1}{C_{\text{DHN}}} &
-\dfrac{1}{C_{\text{DHN}}} &
\dfrac{1}{R_{\text{ext}}C_{\text{DHN}}}
\end{bmatrix}, \\[6pt]
w_{\text{DHN}} &= -\dfrac{A_{\text{coll}} a_1}{C_{\text{DHN}}}\,T_{\text{coll,avg}},
\end{aligned}
\end{equation}
where the second column of $E_{\text{DHN}}$ captures the influence of the ambient temperature $T_{\text{A}}$ on the solar collector,
and $w_{\text{DHN}}$ collects the constant contribution of the collector mean temperature $T_{\text{coll,avg}}$.

To ensure thermally stable and operationally feasible conditions within the DHN, the average water temperature is constrained within admissible bounds:
\begin{equation}
T_{\text{DHN,min}} \le T_{\text{DHN}}(t) \le T_{\text{DHN,max}},
\label{eq:dh_temp_limits}
\end{equation}
where \(T_{\text{DHN,min}}\) and \(T_{\text{DHN,max}}\) denote the lower and upper allowable temperature limits, respectively. These bounds express a physical
operating requirement; in the controller they are enforced as an
exact-penalty soft constraint (Section~\ref{sec:empc_formulation}).

\subsubsection{Electrical Grid Model}
\begin{figure}[ht]
    \centering
    \includegraphics[width=0.9\linewidth]{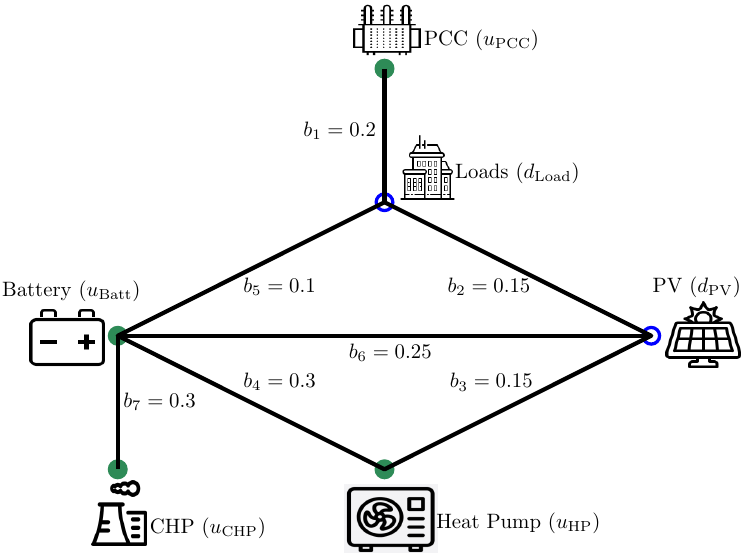}
    \caption{Topology of the electrical network represented as a weighted, undirected graph
    \(\mathcal{G}_e = (\mathcal{N}_e, \mathcal{E}_e, \mathcal{W}_e)\).
    Six nodes, PCC, heat pump, CHP, battery, PV, and load, are interconnected through seven lines, each characterized by a line susceptance \(b_i\).
    Line capacities can be specified to represent physical transfer limits.}
    \label{fig:dc_powerflow}
\end{figure}

Following \cite{purchala_usefulness_2005} and \cite{rose_predictive_2023},
we model the electrical layer as a weighted, undirected graph
\[
\mathcal{G}_e = (\mathcal{N}_e,\,\mathcal{E}_e,\,\mathcal{W}_e),
\]
where $\mathcal{N}_e$ is the set of nodes (buses), $\mathcal{E}_e$ the
set of edges (lines), and $\mathcal{W}_e = \{b_i\}$ the set of edge
weights given by the line susceptances $b_i$.
The topology of the modeled electrical network is illustrated in Fig.~\ref{fig:dc_powerflow}. It consists of six nodes, PCC (Point of common coupling), battery, CHP, heat pump, PV, and loads, interconnected by seven lines, each characterized by a susceptance \(b_i\).

The nodal power injection vector is defined as

\begin{equation}
\begin{aligned}
p_{e,n}(k)
= \big[
&u_{\text{PCC}}(k),\,
-P_{\text{HP,el}}^{\max} \, y_{\text{HP}}(k),\,
P_{\mathrm{av}}(k),\\
&y_{\text{batt}}(k),\,
d_{\text{PV}}(k),\,
-d_{\text{load,el}}(k)
\big]^\top \in \mathbb{R}^{|\mathcal{N}_e|}.
\label{eq:pen_vector}
\end{aligned}
\end{equation}

Here, $u_{\mathrm{PCC}}(k)$ denotes the active power exchange with
the external electrical grid at the PCC (positive~$=$~import,
negative~$=$export) and is itself a control input, because the
imported or exported power is a decision variable in the optimization.
The first four entries of \(p_{e,n}(k)\) correspond to controllable units, and the last two represent exogenous disturbances; the load $d_{\text{load,el}}$ enters as a negative injection.

We assume short, purely inductive lines with constant voltage magnitudes and small voltage angle differences between buses. Under these standard assumptions, the active power flow on each line can be approximated using a linear DC power flow model with
\begin{equation}
    p_{e,e}(k) = \tilde{F}_e\, p_{e,n}(k),
    \label{eq:dcpf_flow}
\end{equation}

where \(p_{e,e}(k)\in\mathbb{R}^{|\mathcal{E}_e|}\) collects the active power flows on all lines at time \(k\), and \(\tilde{F}_e\) is a constant mapping matrix computed from the network’s incidence matrix and the corresponding line parameters (susceptances), providing a linear relationship between nodal injections and line power flows.

A global power balance constraint is imposed to satisfy Kirchhoff’s current law by
\begin{equation}
\mathbf{1}_{|\mathcal{N}_e|}^\top p_{e,n}(k) = 0.
\label{eq:dcpf_balance}
\end{equation}

This nodal active-power balance constraint enforces Kirchhoff's current law
under the DC approximation, and the algebraic DC power flow (DCPF)
formulation introduces no additional dynamic states. For detailed derivations of \(\tilde{F}_e\), see \cite{purchala_usefulness_2005, rose_predictive_2023}.

\subsection{Integrated System Model}
\label{sec:integrated}

Applying zero-order hold (ZOH) discretization at sampling period
$T_s$ to the continuous subsystems (TES, battery, DHN, HP) and
inserting the CHP MLD rows directly yields the unified
discrete-time hybrid model
\begin{equation}
\mathbf{x}^{+} = A_d\,\mathbf{x} + B_d\,\mathbf{u}
+ E_d\,d_{\mathrm{phys}}(k) + w_d,
\label{eq:global_dt}
\end{equation}
with global state and input vectors
\begin{equation}
\begin{split}
\mathbf{x} &=
\begin{bmatrix} x_{\mathrm{TES}} & x_{\mathrm{batt}} & x_{\mathrm{DHN}} & x_{\mathrm{HP}} & x_c & x_\ell \end{bmatrix}^\top, \\
\mathbf{u} &=
\begin{bmatrix} u_{\mathrm{TES}} & P_{\mathrm{ch}} & P_{\mathrm{disch}} & u_{\mathrm{HP}} & \Delta P & \delta^{\mathrm{on,cmd}} & \delta^{\mathrm{start}} \end{bmatrix}^\top,
\end{split}
\label{eq:global_vectors}
\end{equation}
where $x_c = P_{\mathrm{set}}$ and $x_\ell = \delta^{\mathrm{on}}$
are the CHP continuous and binary states, and the physical
disturbance vector is
\[
d_{\mathrm{phys}}(k) =
\begin{bmatrix}
T_{\!A}\;\; G_{\mathrm{coll}}\;\; \dot{Q}_{\mathrm{load}}\;\;
P_{\mathrm{load}}\;\; T_{\mathrm{ext}}\;\; d_{\mathrm{PV}}
\end{bmatrix}^\top.
\]

The state transition matrix is
\footnotesize
\begin{equation}
\renewcommand{\arraystretch}{1.05}
A_d =
\begin{bmatrix}
A_{\mathrm{TES},d} & 0 & 0 & 0 & 0 & 0 \\
0 & 1 & 0 & 0 & 0 & 0 \\
0 & 0 & A_{\mathrm{DHN},d} &
\dfrac{\mathrm{COP}\cdot P_{\mathrm{HP,el}}^{\max}
C_{\mathrm{HP}}}{C_{\mathrm{DHN}}} &
\dfrac{\kappa}{C_{\mathrm{DHN}}} & 0 \\
0 & 0 & 0 & A_{\mathrm{HP},d} & 0 & 0 \\
0 & 0 & 0 & 0 & 1 & 0 \\
0 & 0 & 0 & 0 & 0 & 0
\end{bmatrix}.
\label{eq:Ad_global}
\end{equation}
\normalsize
The diagonal entries of the continuous blocks follow from the
per-block ZOH formula $A_{i,d} = e^{A_{i,c}T_s}$. The battery and
CHP power-setpoint entries equal $1$ because both subsystems are
pure integrators ($A_{i,c}=0$, hence $e^{0}=1$), while the CHP
logical entry is $0$ since $x_\ell^{+}=\delta^{\mathrm{on,cmd}}$
carries no self-dynamics. The off-diagonal entries in the
$x_{\mathrm{DHN}}$ row arise because the HP and CHP thermal
outputs depend on the states $x_{\mathrm{HP}}$ and $x_c$, which enter the DHN dynamics after substituting the HP thermal
relation $P_{\mathrm{HP,th}} = \mathrm{COP}\cdot P_{\mathrm{HP,el}}^{\max}
C_{\mathrm{HP}}\,x_{\mathrm{HP}}$ and the CHP output
map~\eqref{eq:chp_mld_output}. For readability, the off-diagonal
coupling entries in \eqref{eq:Ad_global} and the DHN row of
\eqref{eq:Bd_global} are displayed in their continuous-time form;
the implementation applies exact ZOH discretization
$A_d = e^{A_c T_s}$ to the full coupled system, so all implemented
entries are dimensionally consistent discrete-time maps.

The input matrix is
\footnotesize
\begin{equation}
\renewcommand{\arraystretch}{1.05}
B_d =
\begin{bmatrix}
B_{\mathrm{TES},d} & 0 & 0 & 0 & 0 & 0 & 0 \\
0 & B_{\mathrm{batt},d}^{(1)} & B_{\mathrm{batt},d}^{(2)} &
    0 & 0 & 0 & 0 \\
\dfrac{1}{C_{\mathrm{DHN}}} & 0 & 0 & 0 &
    \dfrac{\kappa(1+\alpha)}{C_{\mathrm{DHN}}} &
    0 &
    \dfrac{\kappa P_{\mathrm{start}}}{C_{\mathrm{DHN}}} \\
0 & 0 & 0 & B_{\mathrm{HP},d} & 0 & 0 & 0 \\
0 & 0 & 0 & 0 & 1 & 0 & 0 \\
0 & 0 & 0 & 0 & 0 & 1 & 0
\end{bmatrix},
\label{eq:Bd_global}
\end{equation}
\normalsize
where $B_{\mathrm{TES},d}$ and $B_{\mathrm{HP},d}$ follow from the
per-block ZOH integral $B_{i,d} = \int_0^{T_s}
e^{A_{i,c}\tau}\,d\tau\,B_{i,c}$, and the battery entries are
$B_{\mathrm{batt},d}^{(1)} = \eta_{\mathrm{ch}}T_s /
(3600\,C_{\mathrm{batt}})$ and $B_{\mathrm{batt},d}^{(2)} =
-T_s / (3600\,\eta_{\mathrm{disch}}\,C_{\mathrm{batt}})$, where the
factor $3600$ converts $T_s$ from seconds to hours consistent with
$C_{\mathrm{batt}}$ in kWh. The CHP rows are inserted directly
from the MLD dynamics~\eqref{eq:chp_mld_dyn} without ZOH
modification, since the CHP is formulated natively in discrete
time at the same $T_s$. The PCC exchange $u_{\mathrm{PCC}}$ does not appear in the
state-space dynamics; it enters only through the DC power-flow
constraints~\eqref{eq:dcpf_balance} below.

The disturbance matrix $E_d$ and affine term $w_d$ are
\[
E_d =
\begin{bmatrix}
E_{\mathrm{TES},d}^{\mathrm{glob}}\\
\mathbf{0}_{1\times 6}\\
E_{\mathrm{DHN},d}^{\mathrm{glob}}\\
\mathbf{0}_{1\times 6}\\
\mathbf{0}_{2\times 6}
\end{bmatrix},
\qquad
w_d =
\begin{bmatrix} w_{\mathrm{TES},d} & 0 & w_{\mathrm{DHN},d} & 0 & 0 & 0 \end{bmatrix}^\top,
\]
where $E_{\mathrm{TES},d}$ and $E_{\mathrm{DHN},d}$ follow from
ZOH. The zero rows for battery, HP, and CHP reflect that these
subsystems receive no direct physical disturbance inputs in their
state equations. The inter-subsystem coupling between the HP and
CHP thermal outputs and the DHN state is enforced through
auxiliary output variables and equality constraints following the
MLD structure of \citep{bemporad_control_1999}.

\paragraph{Algebraic network coupling.}
The state-space dynamics~\eqref{eq:global_dt} govern the temporal
evolution of all component states. At each step $k$, the
electrical outputs must additionally satisfy the DC power-flow
constraints, which enforce physical consistency of nodal
injections and line flows shown in \eqref{eq:pen_vector}, \eqref{eq:dcpf_flow} and \eqref{eq:dcpf_balance}.
Together they
constitute the complete system description: the state-space
equations capture intertemporal storage dynamics, while the
algebraic DC power-flow relations enforce instantaneous
electrical feasibility at every step
\citep{rose_predictive_2023,behrunani_distributed_2024}.

\paragraph{System properties.}
The open-loop system~\eqref{eq:global_dt} is marginally stable:
the battery SOC and CHP power setpoint are pure integrators with
eigenvalues at unity, while all thermal states have eigenvalues
strictly inside the unit disc. Each subsystem is individually controllable from its associated inputs in $B_d$, and the block upper-triangular structure of $A_d$
ensures full state controllability without eigenvalue--input
cancellation across blocks. The presence of binary CHP actuators
does not modify the feasible set structure of the MPC problem:
since the admissible input set $\mathcal{U}$ need not have an
interior for standard MPC feasibility and tracking-stability
results to hold, discrete-valued inputs are accommodated without
further modification \citep{rawlings_risbeck_2017}; the extension
of these results to economic objectives remains an open problem
not addressed here (cf. the discussion of
\mbox{
\citep{risbeck_economic_2020} }\hskip0pt
in Section~\ref{sec:background}, whose
Theorem~1 bound covers the hard-constrained mixed-integer setting,
including the binary CHP commitment variables, via compactness
rather than convexity of the input set). Under the perfect-foresight assumption of
this study, all states are measured directly at each sampling
instant ($C=I$), making the system fully observable; the binary
state $x_\ell$ is measured rather than estimated. The marginal stability of the
integrator states motivates the cyclic-terminal soft constraints
of Section~\ref{sec:empc_formulation}, which anchor
$x_{\mathrm{batt}}$ and $x_c$ to the periodic reference.

\subsection{Cost Function and Optimization Problem}
\label{sec:empc_formulation}
The economic and physical disturbances affecting the system are approximately daily periodic, while still exhibiting short-term fluctuations in loads and electricity prices. To exploit this structure, we compute an economically optimal periodic operating orbit offline, hereafter the \emph{periodic reference}, and use it to anchor the online EMPC during receding-horizon operation. Specifically, the periodic reference provides (i) a periodic target trajectory and (ii) reference terminal values used in softened terminal consistency constraints, whereas the online EMPC continually reoptimizes using the current state and updated forecasts.

Building on the theoretical framework of Section~\ref{sec:background}, the stage cost $\ell(\mathbf{x},\mathbf{u},t)$ is instantiated as a time-varying economic cost driven by the disturbances $d_{\mathrm{phys}}(t)$ and $d_{\mathrm{econ}}(t)$, including the electricity price $c_{\mathrm{el}}(t)$ and the thermal and electrical demand profiles. To improve numerical conditioning and discourage aggressive actuator behavior, we include mild regularization terms on the input magnitude and input increments. Here, $Q_{\mathrm{chp}}(k)$ is the CHP thermal output obtained from the
MLD output map~\eqref{eq:chp_mld_output} and thus depends on both
$\mathbf{x}(k)$ and $\mathbf{u}(k)$; $u_{\text{PCC}}(k)$ denotes the net
active power exchange at the PCC (positive\,$=$\,import).
Furthermore, $c_{\mathrm{gas}}$ is the gas price [€/kg] and
$\mathrm{LHV}$ the fuel's lower heating value, so
$c_{\mathrm{gas}}/(\mathrm{LHV}\,\eta_{\mathrm{th}})$ gives the fuel cost
per unit of thermal output. The electricity price $c_{\mathrm{el}}(t)$ is
a time-varying economic driver.
The instantaneous cost is
\begin{align}
  \ell(\mathbf{x}(k),\mathbf{u}(k),t+k)
  &=
  \frac{c_{\mathrm{gas}}}{\mathrm{LHV}\cdot\eta_{\mathrm{th}}}\,Q_{\mathrm{chp}}(k)\,T_s
  \;+\; c_{\mathrm{su}}\,\delta^{\mathrm{start}}(k) \notag\\[-1pt]
  &\quad
  +\; u_{\mathrm{PCC}}(k)\,c_{\mathrm{el}}(t+k)\,T_s \notag\\[-1pt]
  &\quad
  +\; p_{\text{curtail}}(k)\,c^{+}_{\mathrm{el}}(t+k)\,T_s \notag\\[-1pt]
  &\quad
  +\; \big(P_{\mathrm{ch}}(k) + P_{\mathrm{disch}}(k)\big)\,c_{\mathrm{batt}}\,T_s \notag\\[-1pt]
  &\quad
  +\; W_{T,\mathrm{low}}\,s_{T,\mathrm{low}}(k)
   + W_{T,\mathrm{high}}\,s_{T,\mathrm{high}}(k) \notag\\[-1pt]
  &\quad
  +\; T_s\big(\|R\bar u_k\|_1 + \|L\Delta\bar u_k\|_1\big),
  \label{eq:stage_cost_compact}
\end{align}
where $u_{\mathrm{PCC}}(k)$ is the net grid exchange defined in
Section~\ref{sec:models} (positive~$=$~import), $c_{\mathrm{su}}$ is
the CHP startup wear cost per $0\!\to\!1$ transition [€] and
$\delta^{\mathrm{start}}(k)$ the startup indicator
from~\eqref{eq:global_vectors}, $c^{+}_{\mathrm{el}} =
\max(c_{\mathrm{el}},0)$ is the positive part of the electricity price,
applied to the curtailed PV power $p_{\text{curtail}}(k)$ so that
curtailment is charged at its lost revenue and never credited at negative
prices, $c_{\mathrm{batt}}$ is the battery degradation cost per unit
throughput [€/kWh], and $s_{T,\mathrm{low}},s_{T,\mathrm{high}} \ge 0$ are
the slacks of the DHN temperature bounds~\eqref{eq:dh_temp_limits},
penalized by $W_{T,\mathrm{low}} = W_{T,\mathrm{high}} = 10^{6}$;
$\Delta \mathbf{u}(k)=\mathbf{u}(k)-\mathbf{u}(k-1)$,
in which $\mathbf{u}(k-1)$ is the last applied input: a known
parameter for $k=0$ and a decision variable already present in the
horizon for $k\ge 1$,
and the regularization is applied to component-wise normalized inputs
\begin{equation}
\bar{u}_{i,k} = \frac{u_{i}(k)}{u_{i}^{\mathrm{nom}}},
\qquad
\Delta\bar{u}_{i,k} = \bar{u}_{i,k} - \bar{u}_{i,k-1},
\label{eq:unorm}
\end{equation}
with nominal magnitudes $u_{i}^{\mathrm{nom}}$ from Table~\ref{tab:params_main}.
The battery term penalizes throughput in both directions:
$P_{\mathrm{ch}}$ and $P_{\mathrm{disch}}$ are non-negative and mutually
exclusive by the charge/discharge complementarity binary, so
$P_{\mathrm{ch}}+P_{\mathrm{disch}} = |P_{\mathrm{batt}}|$ exactly and the
term stays linear.

Powers enter \eqref{eq:stage_cost_compact} in kW and prices in
€/kWh, so each power--price product carries an explicit $T_s$ (in hours)
and $\ell$ has units of € per step. The startup and temperature-slack
terms are charged per transition and per violation, respectively, and
therefore carry no $T_s$ factor.
The weights $R = \operatorname{diag}(\lambda_{u,i})$
and $L = \operatorname{diag}(\lambda_{\Delta u,i})$ are diagonal matrices, which are chosen sufficiently small relative to the economic coefficients such that they do not alter the economically optimal behavior and uniform across actuators ($\lambda_{u,i}=10^{-4}$, $\lambda_{\Delta u,i}=1$ for all $i$, except
  the grid exchange, which carries only a magnitude penalty since it is the residual balance variable and smoothing it directly would distort price-following), so as not to materially
  alter the economically optimal behavior: a parameter sweep confirms this configuration increases cost by only $0.10\,\%$ relative to the unregularized baseline while eliminating
  numerical chatter.

To guide the optimizer toward the periodic economic orbit while still permitting controlled deviations, soft terminal constraints are imposed on selected state components. Since the terminal constraints act directly on individual states (TES energy, battery SOC, DHN temperature, HP state, and CHP scheduling variables), the selector matrix $C_T \in \mathbb{R}^{n_s \times n_x}$ introduced in Section~\ref{sec:background} reduces here to rows of the identity matrix (here $C_T$ consists of selected rows of $I$, so $C_T x = x_i$ simply picks the relevant state components). For each selected terminal state $x_{i,N}$ and its corresponding reference value $r_{i,N}$ obtained from the periodic orbit, a slack variable $s_i \ge 0$ bounds the absolute deviation according to
\begin{equation}
\big|\,x_{i,N} - r_{i,N}\,\big| \;\le\; s_i.
\label{eq:terminal_scalar_tube}
\end{equation}
Collecting all terminal slacks in the vector $s_N$ and their corresponding scaling factors in $\sigma>0$, the normalized linear terminal penalty is
\begin{equation}
J_{\mathrm{term}}
=
w_{\mathrm{term}}\,
\mathbf{1}^\top\!\big(\sigma^{-1}\!\odot s_N\big),
\label{eq:terminal_linear_scaled}
\end{equation}
where $\odot$ denotes componentwise multiplication. This corresponds to the terminal penalty $\rho_N^\top s_N$ from Section~\ref{sec:background} with $\rho_N = w_{\mathrm{term}}\,\sigma^{-1}$, where the normalization $\sigma$ ensures comparable numerical scaling across terminal states. The scalar weight $w_{\mathrm{term}}>0$ determines the overall importance of satisfying the terminal tube relative to the stage cost, while $\sigma$ normalizes the individual slacks so that all terminal quantities contribute on comparable numerical levels. The complete finite-horizon cost is
\begin{equation}
J_N(\mathbf{x},t)
=
\sum_{k=0}^{N-1}\ell(\mathbf{x}(k),\mathbf{u}(k),t+k)
+ J_{\mathrm{term}}.
\label{eq:empc_cost_with_terminal}
\end{equation}

\paragraph{Offline periodic reference (target orbit)}
Given an estimate of the disturbance sequence over one period, $\{d_k\}_{k=0}^{N_p-1}$, we compute offline a feasible periodic state--input trajectory $\{x_k^{\mathrm{ref}},u_k^{\mathrm{ref}}\}_{k=0}^{N_p-1}$ by solving
\begin{align}
\{x_k^{\mathrm{ref}},u_k^{\mathrm{ref}}\}_{k=0}^{N_p-1}
\in \arg\min_{\{x_k,u_k\}} \quad &
\sum_{k=0}^{N_p-1} \ell(x_k,u_k,k)
\label{eq:offline_ref}\\
\text{s.t.}\quad
x_{k+1} &= A_d x_k + B_d u_k + E_d d_k + w_d, \nonumber\\
(x_k,u_k) &\in \mathcal{X}\times\mathcal{U}, \quad \mathrm{DCPF}(x_k,u_k), \nonumber\\
x_{N_p} &= x_0. \nonumber
\end{align}
The periodicity constraint $x_{N_p}=x_0$ enforces a closed orbit consistent with the assumed periodic disturbance pattern. Here $x_0$ (equivalently $x_{N_p}$) is itself an optimization variable: the orbit's closure point is free, and only periodicity, not a prescribed terminal state, is imposed. The resulting optimizer defines the periodic targets $\{x_k^{\mathrm{ref}},u_k^{\mathrm{ref}}\}$ and the corresponding terminal reference values $r_{i,N}$ used in~\eqref{eq:terminal_scalar_tube}.

\paragraph{Online EMPC (receding-horizon operation).}
During operation, at each time $t$ the controller solves a finite-horizon EMPC problem initialized at the measured state and driven by updated forecasts, minimizing $J_N(\mathbf{x},t)$ in~\eqref{eq:empc_cost_with_terminal}. The problem is re-solved at every sampling instant $t$ (period $T_s$): the horizon terms use the time-shifted forecasts $d_{\mathrm{phys}}(t{+}k)$ and stage costs $\ell(\cdot,t{+}k)$, and the terminal reference $r_N=C_T\,x^{\mathrm{ref}}_{t+N}$ is obtained by indexing the \emph{fixed} periodic reference at $t{+}N$. The periodic reference is computed once before operation and is never recomputed during the run, so the terminal reference shifts smoothly along it as $t$ advances rather than jumping between successive solves. The periodic reference anchors the online optimization to a feasible economic operating regime, while the slack variables $s_N$ allow temporary deviations when required by forecast changes or constraint activity.

\subsection{Experimental Setup}
\label{sec:setup}
The simulation setup is based on the multi-energy infrastructure of the University of Stuttgart Campus Vaihingen,
which includes a CHP plant, a TES system,
PV generation, and both thermal and electrical distribution networks.
In its planned future configuration, the campus will additionally operate a large-scale heat pump
using waste heat from the local high-performance computing center \citep{badenwuerttemberg_start_2025}.
For the present study, this real infrastructure is augmented with a solar-thermal collector field and a battery system
to explore the full potential of coordinated electro-thermal operation.
Heat and electricity are coupled in the resulting test system through the
CHP, the heat pump, and the shared campus connection point.
The overall system topology is illustrated in Fig.~\ref{fig:overview}.

All internal computations use SI base units (J, W, K), consistent
with the continuous-time and discretized state-space models;
Table~\ref{tab:params_main} reports values in common engineering
units (kW, MW, kWh, $^\circ$C) for readability. Thermal and electrical demand, PV irradiance, and ambient temperature follow
representative measured profiles. A periodic economic reference trajectory is
computed offline over multiple days and used as the terminal target for the EMPC controller.

The electricity price series is composed of German
intraday prices from 2024, mapped onto the 2019 timebase of the
measured campus demand and weather profiles; the gas price is held
constant (Table~\ref{tab:params_main}). This splice preserves the
realistic short-term electricity-price volatility relevant to
storage and CHP dispatch, at the cost of severing the physical
correlation between weather and price; this is disclosed as a
limitation.

Table~\ref{tab:params_main} summarizes the key parameters of the case study, including network
limits, component ratings, and economic data. The MPC sampling time is set to $T_s = 300\,\mathrm{s}$,
and the prediction horizon spans 24 hours and therefore $N = 288$ steps. SOC limits of $0.2$–$0.9$ ensure
practical depth-of-discharge operation.
For the EMPC formulation, a uniform regularization scheme is applied to normalized
actuator signals: a small magnitude penalty $\lambda_{u} = 10^{-4}$
improves numerical conditioning, and a smoothness penalty
$\lambda_{\Delta u} = 1$ is applied uniformly across all continuous
inputs, including the battery and the TES.

\begin{table}[t]
\centering
\caption{Key parameters of the EMPC case study.}
\scriptsize
\label{tab:params_main}
\begin{tabular}{lll}
\toprule
Category & Parameter & Value \\
\midrule
\textit{MPC}
& Sampling time $T_s$ & $300$\,s \\
& Horizon $N$ & $288$ steps $= 24$\,h \\
& Terminal weight $w_{\mathrm{term}}$ & $10^3$ \\
\midrule
\textit{Regularization}
& All actuators (grid: magnitude only) & $10^{-4}$ / $1$ \\
\midrule
\textit{Costs / Efficiencies}
& LHV & $12\,500$\,Wh/kg \\
& Gas price $c_{\mathrm{gas}}$ & 0.625 €/kg \\
& CHP $\eta_{\rm th}/\eta_{\rm el}$ & $0.58$ / $0.24$ \\
& Heat pump COP & $3.252$ \\
\midrule
\textit{Heating grid / TES}
& $1/R_{\mathrm{ext}}$ (grid conductance) & $183$\,W/K \\
& DHN thermal capacitance $C_{\mathrm{DHN}}$ & $6.2\times10^{9}$\,J/K \\
& TES $UA$ & $500$\,W/K \\
& TES energy $Q_{\mathrm{TES}}^{\max}$ & $2.7\times10^{9}$\,J \\
& High/Low TES temperatures $T_H/T_L$ & $100/70\,^{\circ}$C \\
& TES power $P_{\mathrm{TES}}^{\max}$ & $500$\,kW \\
\midrule
\textit{Generation}
& CHP $P_{\mathrm{CHP}}^{\min}/P_{\mathrm{CHP}}^{\max}$
                                          & $10$\,kW / $1.0$\,MW \\
& CHP ramp $\Delta P$ & $\pm60$\,kW/step \\
& CHP startup $\Delta P_{\mathrm{start}}$ & $15$\,kW \\
& HP $P_{\mathrm{el}}^{\max}$ & $2$\,MW \\
& Solar coll.\ $A_{\mathrm{coll}}$ & $4500$\,m$^2$ \\
& Coll.\ $\eta_0 / a_1$
  & $0.76$\,[-] / $1.82$\,W/(m$^2$K) \\
& PV $P_{\mathrm{PV}}^{\max}$ & $10$\,MW \\
\midrule
\textit{DC power flow}
& Nodes / lines & $6$ / $7$ \\
& PCC limit $P_{\mathrm{PCC}}^{\max}$ & $8.0$\,MW \\
& Line limit $P_{\ell}^{\max}$ & $10$\,MW \\
& Line susceptance $b_\ell$ & $0.10$--$0.30$ \\
\midrule
\textit{Battery}
& Capacity $C_{\mathrm{batt}}$ & $5.0\times10^{7}$\,Wh \\
& Power $P_{\mathrm{batt}}^{\max}$ & $2.0$\,MW \\
& SOC bounds & $0.2$--$0.9$ \\
& Round-trip efficiency $\eta_{\mathrm{batt}}$ & $0.95$ \\
& Throughput cost & $10$\,€/MWh \\
\bottomrule
\end{tabular}
\end{table}

\section{Results}
\label{sec:results}

The following results evaluate the closed-loop behavior of the proposed cyclic-terminal EMPC in three parts: (i) a baseline scenario establishing the qualitative behavior of one closed-loop trajectory under perfect foresight, (ii) the orbit-alignment reading of a joint sweep over the prediction horizon $N$ and the terminal weight $w_{\mathrm{term}}$, and (iii) the closed-loop-cost reading of that same joint sweep. Parts (ii) and (iii) are two views of one $14 \times 8$ grid of $112$ closed-loop runs rather than independent one-dimensional sweeps, because horizon length and terminal weight are not separable design choices: each substitutes partially for the other, and that substitution is itself the central result reported below. All online simulations cover three consecutive days with sampling time $T_s=300$\,s (5\,min) and use perfect forecasts for exogenous signals. The initial condition $x_0$ is set to $x^{\mathrm{ref}}_0$ of the periodic reference, so that the reported gaps isolate finite-horizon effects rather than transient response to $x_0 \neq x_0^{\mathrm{ref}}$.

\subsection{Baseline Scenario}
\label{sec:baseline}

The baseline case characterizes the closed-loop behavior of the
proposed cyclic-terminal EMPC under idealized conditions with
perfect forecasts of all exogenous signals and a simulated plant
matching the controller model. The results therefore isolate the
behavior induced by the economic objective, the hybrid feasibility
constraints, and the cyclic-terminal structure.

The periodic reference $\{x_k^{\mathrm{ref}},u_k^{\mathrm{ref}}\}$
is computed offline as one cyclic problem over eight consecutive
days ($2304$ steps), with the periodicity constraint imposed
between the first and the last step of that window. The
closed-loop simulation then covers the first three days of the
same window. The window must exceed those three days by at least
the longest online prediction horizon tested ($N=864$, i.e.\
$72$\,h), since at every step of the closed-loop simulation the
controller reads reference states and inputs up to $N$ steps
beyond the current time; eight days leave five days of reference
beyond the simulated window, two more than the longest horizon
requires. Online, we use a receding-horizon EMPC with a 24\,h
prediction horizon of $N=288$ steps and terminal penalty weight
$w_{\mathrm{term}}=10^3$, reoptimizing at each step. Of the tested weight grid $\{0, 30, 100, 200, 300, 10^3, 10^4,
10^6\}$, of which the seven weights up to $10^4$ are shown in the figures
(Section~\ref{sec:terminal_sweep}), the battery state-of-charge alignment becomes exact
between $w_{\mathrm{term}}=200$ and $w_{\mathrm{term}}=300$; the
chosen $w_{\mathrm{term}}=10^3$ sits roughly half a decade above
this activation edge, providing margin without pushing into the
extreme-weight regime (Section~\ref{sec:terminal_sweep}). Measured
by the turnpike fraction of Section~\ref{sec:grid_metrics}, the
battery is the only coordinate that reaches exactly $1.000$ at
every horizon: the DHN temperature plateaus at a mean of $0.96$
without attaining unity across all horizons at any single weight,
and the TES, heat-pump, and CHP fractions stay below unity at every
weight tested. Measured by the terminal deviation instead, the TES
and CHP terminal states are pinned to the reference to machine
precision from the smallest nonzero weight on; the two metrics
answer different questions and should not be conflated. The
baseline therefore demonstrates the paper's own recommended
operating point rather than an arbitrarily strong
setting; the cost--alignment trade-off across weights is analyzed
in Section~\ref{sec:terminal_sweep}.

\begin{figure}[h!]
    \centering
    \includegraphics[width=\linewidth]{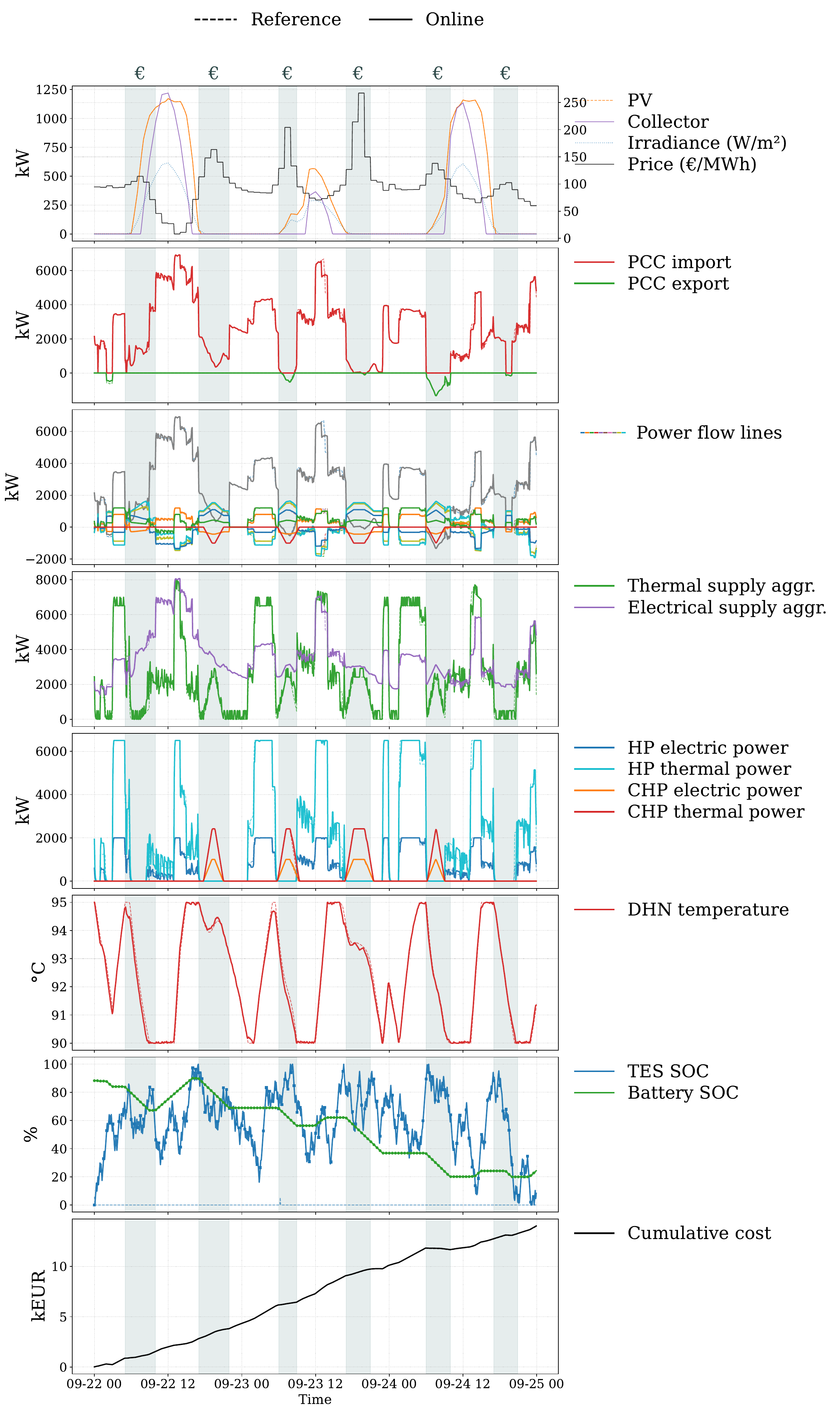}
    \caption{Baseline EMPC operation under perfect forecasts with
    a 24\,h prediction horizon. Panels show: (i) exogenous signals
    (PV irradiance, solar-collector input, electricity price),
    (ii) PCC import/export, (iii) DC power flow lines,
    (iv) aggregated thermal and electrical supply,
    (v) heat-pump and CHP dispatch, (vi) DHN temperature,
    (vii) TES energy and battery SOC, and (viii) cumulative stage
    cost. Dashed lines: offline periodic reference; solid lines:
    online EMPC.}
    \label{fig:system_overview_baseline}
\end{figure}

Figure~\ref{fig:system_overview_baseline} summarizes the coupled
thermal-electrical operation. The online EMPC tracks the offline
periodic orbit nearly identically for the continuous network states
and respects all thermal and electrical feasibility limits.

The flexibility allocation is clearly asymmetric across
energy carriers, but the reverse of a thermal-dominates-medium-term
picture. The battery performs slow, multi-hour price arbitrage: over
the three-day window it completes a small number of deep sweeps across
its state-of-charge range ( 7 sign reversals total, correlation
with price $-0.77$ for net charging power $P_{\text{ch}}-P_{\text{disch}}$,
positive in charging) \footnote{All correlations reported in this section are
Pearson coefficients computed between the closed-loop signal and the
intraday electricity price $c_{\mathrm{el}}(t)$, evaluated
pointwise over all $864$ samples of the three-day window at the
$T_s=300$\,s simulation resolution, with no lagging, smoothing, or
resampling.}, charging during cheap-price intervals and
discharging into expensive ones; it, not the thermal subsystem, is
the medium-term mover. The TES, by contrast, is a fast,
near-continuous cycler ( $\approx 410$ sign reversals, sub-hourly
period) whose heat-flow signal is not itself strongly price-driven
(correlation of the TES heat-flow signal $Q_{\mathrm{hs}}$ with
price is $+0.03$, near zero, with the sign a function of the
discharge-positive flow convention used here; note that the TES
\emph{level} state $x_0$ itself correlates more appreciably with
price, at $+0.35$ ) but instead integrates a persistent heat-pump overproduction
bias: the heat pump supplies on average $0.5\,\%$ more thermal
output than the periodic reference ( $+10.8$ \,kW), and the TES
absorbs this excess. This is not a contradiction of the heat pump's
own price-driven dispatch: what reaches the TES is not that
price-driven signal directly but the residual mismatch between
heat-pump output and instantaneous demand, i.e.\ the overproduction
bias, which is why the TES itself shows negligible price correlation
even though its charging source is price-driven upstream. The
periodic reference keeps the TES nearly
inert, whereas the online EMPC actively drives it far across its
range; the mechanism behind this receding-horizon drift is analyzed
in Section~\ref{sec:discussion}. Rather than following a fixed
heuristic, the controller alternates between direct heat-pump supply
and storage exchange depending on the active thermal constraints and
the current electrical operating point.

On the electrical side, the CHP is dispatched predominantly during
the highest-price portion of the window ( $35\,\%$ duty on the
second, highest-price day versus $14\,\%$ and $12\,\%$ on the
first and third days; correlation of the commitment signal with price $0.70$ ) rather than
being spread evenly across all higher-price periods; DHN and TES
bounds are occasionally active during this period but do not
materially constrain the CHP schedule. In contrast to the TES, the
battery tracks the periodic reference tightly, executing the same
slow price-arbitrage sweep described above. Grid exchange at the
PCC follows the same economic logic but remains constrained by
local demand, unit ratings, and PCC capacity.

Constraint activity explains the remaining deviations from a
price-only strategy: DHN and PCC bounds bind during the sharpest
price swings. PV curtailment is the remaining relief mechanism
whenever additional injection would violate network or storage
limits, but it is not exercised in this window: the curtailment
variable $p_{\text{curtail}}$ is identically zero over all three
days, so the local PV is absorbed in full here. Curtailment
does become active in the reproduction week used in
Section~\ref{sec:grid_aac}.

Panel~(viii) of Figure~\ref{fig:system_overview_baseline} shows
the cumulative stage cost over the three simulated days.
The online EMPC and the offline periodic reference track
each other closely: the online total cost exceeds the reference by
only $0.77\,\%$ over the three-day window. A
finite-horizon, cyclic-terminal-anchored controller therefore stays
within roughly one percent of the jointly-optimized offline
reference on this window; this small, near-vanishing gap is
consistent with, and not a counterexample to, the asymptotic
average-performance bound of Theorem~1 in \citep{risbeck_economic_2020}:
over a quasi-periodic exogenous signal the receding-horizon cost is
expected to track the periodic-orbit cost closely, with the residual
gap attributable to the finite-horizon effects analyzed in detail in
Section~\ref{sec:grid_aac}. Part of this residual gap is a storage-inventory
bookkeeping effect rather than a true economic loss: the online
terminal TES level exceeds the reference's own by $\approx
62$\,kWh$_{\mathrm{th}}$ ($\approx 224$\,MJ), worth well under
$0.1\,\%$ of total cost at window prices, so the reported gap is a
conservative upper bound.

\subsection{\texorpdfstring{The Joint Horizon--Weight Grid: Orbit Alignment and Closed-Loop Cost}{The Joint Horizon-Weight Grid: Orbit Alignment and Closed-Loop Cost}}
\label{sec:grid}
\label{sec:grid_turnpike}
\label{sec:horizon_sweep}
\label{sec:grid_aac}
\label{sec:terminal_sweep}

\begin{figure}[ht!]
    \centering
    \includegraphics[width=0.9\linewidth]{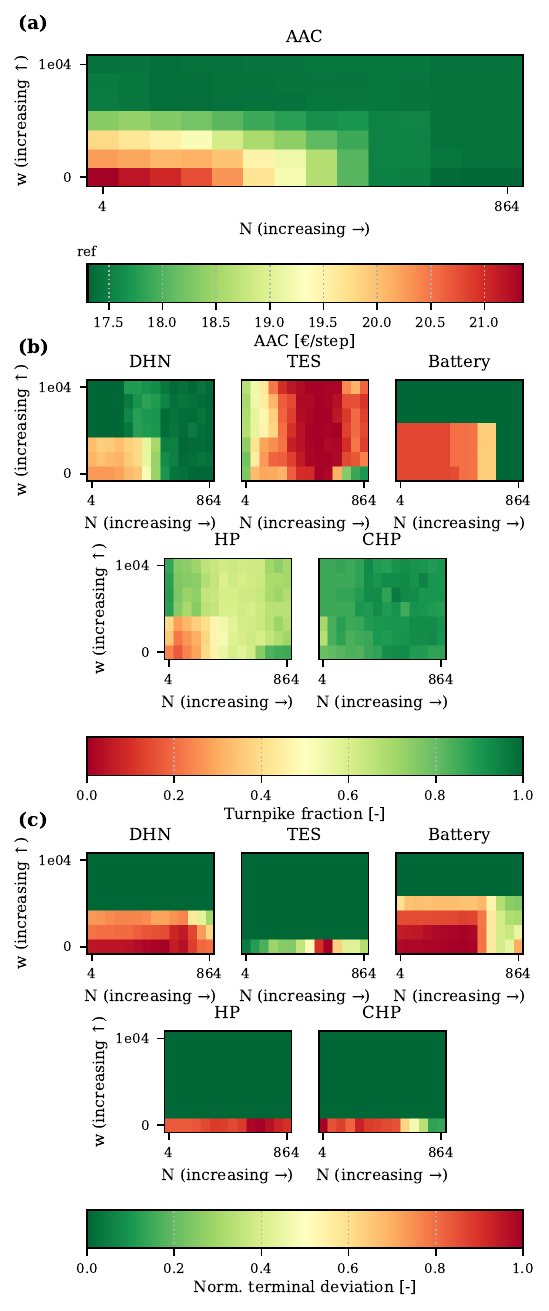}
    \caption{Joint horizon--weight grid, price week 2019-09-22,
    $14 \times 8$ campaign of $N$ (columns) and $w_{\mathrm{term}}$ (rows),
    $112$ runs; all numbers quoted in the text are this campaign's own. The
    panels display the seven weights up to $10^4$: $w_{\mathrm{term}}=10^6$
    behaves like $10^4$ throughout, matching it exactly in the battery
    coordinate, to within $0.31\,\%$ in AAC and to within $0.07$ in the
    remaining turnpike fractions except the unanchored TES, and is omitted for
    legibility. All quantitative statements in the text are computed over the
    full eight-weight campaign.
    (a) AAC, green at/below reference, red above.
    (b) Turnpike fraction per state, green full alignment, red none.
    (c) Normalized terminal deviation per state, $[0,1]$, green tight to
    reference. Same activation edge in all panels; TES is the one state
    neither parameter anchors.}
    \label{fig:grid_second_week}
\end{figure}

Both readings that follow draw on a single campaign: a full
factorial sweep of the prediction horizon over the fourteen-point grid
$N \in \{4,8,12,16,24,36,48,72,144,288,432,576,720,864\}$, spanning $20$\,min to
$72$\,h of lookahead, against the terminal weight
$w_{\mathrm{term}} \in \{0,\, 30,\, 10^2,\, 2\cdot10^2,\, 3\cdot10^2,\, 10^3,\,
10^4,\, 10^6\}$, i.e.\ $112$ closed-loop runs. All runs share the same
three-day window, the matched initial condition $x_0 = x_0^{\mathrm{ref}}$, and
an otherwise identical configuration. The
weight grid is deliberately dense between $10^2$ and $10^3$, which is where
the terminal tube activates. Setting $w_{\mathrm{term}}=0$ removes the soft
terminal constraint~\eqref{eq:tube_final} entirely and reduces the scheme to a
plain economic MPC with no anchoring to the periodic orbit, so the bottom row
of each grid is the unanchored baseline against which the terminal condition
is assessed. The two metrics used to read the grid are the turnpike fraction
(state alignment) and the AAC ( closed-loop cost ), defined in
Section~\ref{sec:grid_metrics} below; the alignment reading is discussed
first, the cost reading second. A third quantity, the normalized terminal
deviation per state, is reported alongside them in
Figure~\ref{fig:grid_second_week}(c) as a direct check that the terminal tube
is doing what the turnpike fraction attributes to it; it is corroborating
rather than a separate reading.

\label{sec:grid_metrics}
The average closed-loop cost AAC and the turnpike fraction are
defined next, together with the methodological caveats that apply
to both readings. Theorem~1 in \mbox{
\citep{risbeck_economic_2020}
}\hskip0pt
guarantees that under the hard terminal conditions recalled in
Section~\ref{sec:background}, the asymptotic average closed-loop
cost satisfies
\begin{equation}
\limsup_{T \to \infty} \frac{1}{T} \sum_{k=0}^{T-1}
\left[\ell(x(k),u(k),k) - \ell(x^{\mathrm{ref}}(k),
u^{\mathrm{ref}}(k),k)\right] \leq 0.
\end{equation}
We report the empirical average stage cost over the 72\,h
simulation window, denoted AAC, as a finite-window approximation
of this asymptotic quantity (the terminal condition here is a soft
penalized tube, tight to solver tolerance for $w_{\mathrm{term}}
\geq 10^4$; Section~\ref{sec:background} states the soft-vs-hard
distinction).

The turnpike fraction, defined for each state $i$ as the fraction of
closed-loop time steps at which the normalized absolute deviation
satisfies
\begin{equation}
\frac{|x_i(t) - x_i^{\mathrm{ref}}(t)|}{\sigma_i} < 1,
\label{eq:turnpike_def}
\end{equation}
where $\sigma_i$ is a state-specific physical tolerance:
$\sigma_{T_{\mathrm{grid}}} = 0.4$\,°C,
$\sigma_{E_{\mathrm{TES}}} = 1.35\times10^{8}$\,J,
$\sigma_{\mathrm{SOC}} = 0.025$,
$\sigma_{x_{\mathrm{HP}}} = 5$\,[-], and
$\sigma_{P_{\mathrm{CHP}}} = 50$\,kW.
Here $x_{\mathrm{HP}}$ denotes the heat pump's internal
(unnormalized) dynamic state, not the normalized input
$u_{\mathrm{HP}} \in [0,1]$; the tolerance $\sigma_{x_{\mathrm{HP}}}$
is expressed in those internal state units. Note that this
fraction is computed on the realized \emph{closed-loop} trajectory
$x_i(t)$, not on the open-loop predicted trajectory within each
solved horizon; it is therefore best read as a closed-loop
alignment/tracking diagnostic inspired by turnpike terminology,
rather than as a direct measurement of the open-loop turnpike
property that the underlying theory concerns. Two disclosures
about these tolerances. First, the battery result reported below is obtained with
the tightened battery tolerance $\sigma_{\mathrm{SOC}} = 0.025$, i.e.\ the
battery tracks the periodic reference within $2.5\,\%$ SOC throughout.
Second, these measurement tolerances are diagnostic quantities chosen for
reporting and are distinct from the terminal-tube normalisation constants
used inside the controller, which are
$3.0$\,°C, $1.348\times10^{8}$\,J, $0.05$, $5$\,[-] and $100$\,kW in the same
order; the two sets should not be conflated.

The turnpike fraction of~\eqref{eq:turnpike_def} is evaluated for each
of the five state groups over the grid shown in
Figure~\ref{fig:grid_second_week}(b).

\paragraph*{Orbit alignment across the grid.}
Two
distinct alignment mechanisms are visible, and they act on different axes. The
DHN temperature and the battery SOC are governed almost entirely by the terminal
weight: the DHN fraction rises from well below half at the shortest horizons for
low weights to at least $0.80$ at \emph{every} horizon once
$w_{\mathrm{term}} \geq 2\cdot10^2$, and the battery SOC fraction stays at or
below $0.363$ at every horizon shorter than $N=576$ for all weights up to and
including $2\cdot10^2$, and is exactly $1.000$ in all $56$ cells
with $w_{\mathrm{term}} \geq 3\cdot10^2$, independently of $N$. On this
primary week the $w_{\mathrm{term}} = 2\cdot10^2$ row is indistinguishable
from the unanchored rows in the battery coordinate, whereas on the secondary
week it is genuinely intermediate (horizon mean $0.569$ against $0.137$ at
$w_{\mathrm{term}}=0$, the latter carried entirely by the three longest
horizons, since the unanchored fraction stays below $0.01$ for all
$N \leq 432$), so the edge is sampled from below on one week and
straddled on the other; the tube
therefore anchors the battery through the interior of the horizon, not only at
the terminal step, and there is no longer a staged, state-by-state convergence
ordering once the edge is crossed. Terminal-tube
activation therefore appears in the alignment grid as a sharp horizontal edge in
$w_{\mathrm{term}}$, located in $(10^2,\, 3\cdot10^2]$, a step in the
grid resolution tested, sampled directly at $2\cdot10^2$ and $3\cdot10^2$,
and not as a diagonal
trade-off against horizon length.

Lookahead is nevertheless a partial substitute for anchoring, which is
the interaction that motivates reading the sweep as a grid rather than as two
one-dimensional cuts. In the unanchored bottom row the battery SOC fraction
recovers on its own well before the longest tested horizon: on this primary
price week (2019-09-22) it is $0.363$ at $N=432$ and reaches $1.000$ at
$N=576$, i.e.\ two diurnal cycles of lookahead, so a
plain economic MPC given two to three diurnal cycles of lookahead attains the
alignment that the terminal condition already delivers at $N=4$, i.e.\ $20$\,min
of lookahead. The DHN state shows the same substitution but stays
partially anchored throughout: its unanchored fraction has a minimum of
$0.269$ and never falls below $0.2$ on this week, climbing toward unity as
the horizon lengthens. The terminal
condition thus buys, at negligible
online cost, alignment that would otherwise require a two-order-of-magnitude
longer horizon.

The TES is the exception on both axes: no weight anchors it, but its
sensitivity to $w_{\mathrm{term}}$ is itself strongly horizon-dependent
rather than uniformly weak. The max/min ratio of its fraction across the
eight weights at a fixed horizon stays under a factor of two only for the
shortest horizons tested ($N \leq 12$). Beyond that the ratio is large but not
monotone in $N$, spanning $4.5\times$ at $N=288$ and $268\times$ at $N=432$ on
this primary week (2019-09-22), and it is unbounded at $N=144$, where the
smallest fraction over the eight weights is zero; at $N=432$ the secondary
week gives $\approx 64\times$. It is also non-monotone in $N$: large at the shortest
horizons, falling over the mid-range of horizons, and rising again for
$N \geq 576$. The CHP fraction is by contrast nearly invariant over the whole
grid, and well below unity, capped by the quantized nature of its dispatch
rather than by either design parameter: its row means span only
$0.864$ to $0.897$ across all eight weights on this week. The heat pump sits
between the two behaviors. It responds to the same terminal-weight edge as the
DHN and the battery, with row means rising from $0.498$--$0.528$ for
$w_{\mathrm{term}} \leq 10^2$ to $0.670$--$0.684$ for
$w_{\mathrm{term}} \geq 2\cdot10^2$ and its grid-wide minimum lifting from
$0.170$ to $0.587$, but it plateaus far below unity and retains only a mild
horizon dependence above the edge. The TES mechanism is
analyzed in Section~\ref{sec:discussion}.

Figure~\ref{fig:grid_second_week} is the paper's only illustration of
the grid campaign: it shows the two readings, alignment and closed-loop
cost, together with the terminal-deviation check, over the full $14 \times 8$
grid, and is therefore the
visual reference for both readings that follow; all numbers quoted below are
this campaign's own. The full campaign was repeated on a secondary price
week (2019-09-14) as a reproduction check and gave similar results
throughout, confirming that the lookahead a plain economic MPC needs
is price-week dependent, which is exactly the dependence the terminal
condition removes.
  Read over the grid, the Theorem~1 bound (Section~\ref{sec:grid_metrics}) is
attained tightly wherever the terminal condition is active and only at long
horizons where it is not; the cost reading below quantifies that substitution
in cost terms.
The remainder of this subsection collects the integer-dispatch and
solve-time diagnostics of the $w_{\mathrm{term}}=10^3$ row (the baseline
weight, Section~\ref{sec:baseline}), which the alignment grid of
Figure~\ref{fig:grid_second_week}(b) does not display.
The discrete CHP/HP commitment schedule is
  essentially frozen for $N \geq 144$ (8 switches, 4 startups over the three-day
window; 10 switches, 5 startups at $N=288$ ), while at $N=72$ the
same cost is instead achieved with markedly more switching ( 58
switches, 29 startups), i.e.\ more myopic dispatch that is
cost-neutral but not schedule-neutral.
With the terminal condition active, cost is therefore saturated in $N$
across the whole grid row, while the mean solve time grows roughly
linearly from $275$ \,ms ($N=72$) to $3379$ \,ms ($N=864$): a factor of
  $\approx 12.3\times$ against a $12\times$ increase in $N$, i.e.\ an
empirical scaling exponent of $\approx 1.01$, consistent with linear
rather than superlinear growth in the horizon, with the growth in fact
flattening slightly between $N=720$ and $N=864$. The Theorem~1 average-performance
bound of \citep{risbeck_economic_2020} is thus attained essentially
tightly at every tested horizon at this weight, although the formal
bound is not directly in force for the implemented soft-penalty
scheme (Section~\ref{sec:methodology}). The raw gaps
are moreover conservative upper bounds: the closed loop ends the finite
window holding up to $\approx 2.3$ \,GJ of TES inventory relative to
the reference terminal state, peaking at $N=72$, with several cells
(including $N=4$ and $N=12$) essentially at zero; the baseline cell
($N=288$) carries $\approx 0.22$\,GJ. This uncredited inventory is
not counted by the finite-window metric; the largest gap ($N=72$)
coincides with the largest un-credited inventory.
Since the CHP schedule is essentially frozen across $N \geq 144$, this
residual is not CHP-switching noise; it reflects sub-integer variation in
continuous dispatch and in HP switching.

  The turnpike metric
validates the controller on all states except TES, which carries no stage
cost and so is not anchored by the running cost away from the terminal
step (Section~\ref{sec:discussion}).
TES deviations are bimodal: at any given time the TES state is
either on the reference or far off it, so the fraction is insensitive to
the precise tolerance choice.

The CHP dispatch likewise agrees with the reference: over the
same three-day window, the online controller and the offline periodic
reference dispatch the CHP at closely agreeing mean power and duty
cycle, with no systematic under-commitment.

  The alignment reading therefore locates a single sharp edge in
$w_{\mathrm{term}}$ and shows lookahead substituting for anchoring only at
horizons two orders of magnitude longer. Whether that alignment is bought at
an economic price, or is simply free, is not answerable from the turnpike
fractions alone: the same grid must be read a second time in closed-loop cost.

\paragraph*{Closed-loop cost across the grid.}

The closed-loop AAC is evaluated over the same $14 \times 8$ grid,
diverging about the window-matched periodic reference of $17.2978$ \,€/step;
Figure~\ref{fig:grid_second_week}(a) shows the corresponding cost surface. The grid
separates the two design parameters cleanly. Along
the horizon axis, cost is strongly horizon-dependent only while the terminal
condition is inactive: at $w_{\mathrm{term}}=0$ the AAC falls monotonically
with lookahead, from $21.359$ \,€/step at $N=4$ ( $+23.5\,\%$ relative to the
reference) to $17.3138$\,€/step at $N=864$ ($+0.09\,\%$). That row spread
contracts by an order of
magnitude as the weight is raised across the activation edge identified in
the alignment grid, the same contraction visible in
Figure~\ref{fig:grid_second_week}(a). Above the edge, all $56$ cells with
$w_{\mathrm{term}} \geq 3\cdot10^2$ lie within $+0.39\,\%$ to $+1.02\,\%$ of
the reference, at every horizon from $20$\,min to $72$\,h.

The same substitution is visible along the other axis: the spread over
the weight column contracts by more than an order of magnitude between the
shortest and longest horizons, i.e.\ with two to three diurnal cycles of
lookahead (two on the primary week, three on the secondary) the
terminal weight barely matters, and with the terminal weight active the horizon
barely matters. Within the anchored region the residual variation is
non-monotone in $N$ and of the order of one per cent; we attribute it to
integer-switching noise rather than to a systematic horizon effect, and
consequently we do not read longer lookahead as either beneficial or harmful
once the terminal condition is active. In practice the cyclic terminal condition removes the horizon from the
performance trade-off: $N$ can then be chosen on solve-time grounds alone.

  No cell in the grid prices out below the reference; the lowest AAC in the
campaign is $17.3138$ \,€/step ($+0.09\,\%$) at $(N,w_{\mathrm{term}}) =
(864,0)$, i.e.\ the longest-horizon, zero-weight cell, consistent with the
alignment reading that sufficient lookahead alone can reach the orbit
without the terminal penalty. This cell is not an outlier: it is the same
horizon-substitutes-for-anchoring effect read in cost rather than in
turnpike fraction.

The baseline of Section~\ref{sec:baseline} is itself a grid cell,
$(N,w_{\mathrm{term}}) = (288,\,10^3)$, at $17.4303$ \,€/step, $+0.766\,\%$
from the reference.

  A finer one-dimensional weight sweep at $N=288$ (price week 2019-09-14)
gave similar results to the coarser grid: the same activation edge and
plateau, the terminal SOC deviation collapsing $\approx 100\times$ across
the transition, and no change in the discrete CHP dispatch pattern.

  Closed-loop behavior is invariant over the full plateau
$w_{\mathrm{term}} \in [10^3, 10^{15}]$ tested here. We nonetheless
recommend $w_{\mathrm{term}} \in [10^3, 10^{12}]$ and
$w_{\mathrm{term}} = 10^3$ as the default: the lower end sits
comfortably above the activation edge, while the upper cap retains
three decades of margin against the ill-conditioning that a terminal
penalty many orders of magnitude above the stage cost would
eventually induce in the MILP relaxation, although no conditioning
failure was observed at any weight tested.

The \emph{predicted} CHP terminal state is anchored to
the periodic reference for $w_{\mathrm{term}} \geq 10^4$ in the same
way as the continuous states: both the predicted CHP power setpoint
$P_{\mathrm{set}}(N)$ and the binary on/off state
$\delta^{\mathrm{on}}(N)$ at the horizon end match
$x_{t+N}^{\mathrm{ref}}$ to within solver tolerance. This concerns
the open-loop terminal prediction only; as noted above, the
closed-loop CHP trajectory does not align exactly with the reference
at any tested weight.

\subsection{Discussion}
\label{sec:discussion}

The horizon and terminal sensitivity analyses together characterize the closed-loop performance of the proposed cyclic-terminal EMPC. Four points follow from the grid.

First, the prediction horizon and the cyclic terminal condition act as
substitutes rather than as independent tuning knobs. Without anchoring, the
controller must look ahead far enough to discover the periodic orbit on its own,
and the cost gap closes only once the horizon spans roughly two to three
diurnal cycles (two on the primary week, three on the secondary).
The terminal condition supplies that information directly, so the same
performance is reached with a horizon two orders of magnitude shorter. This is
the practically relevant statement for deployment: horizon length can be traded
against terminal-condition design, and the computational cost of the former is
what makes the trade worth making, since the mean solve time grows roughly
linearly in $N$ while the achieved cost does not improve.

  Second, this saturation in $N$
(Section~\ref{sec:horizon_sweep}) is why $N=288$ is the concrete design
recommendation for deployment: it matches the cost of shorter horizons
while retaining the low-switching dispatch pattern and spanning a full
diurnal cycle of the exogenous signals. This horizon insensitivity is
not incidental but the expected regime given the structure of the
optimal periodic reference. The reference is quasi-static: its TES
level stays near-constant and nearly empty and all storage
cycling is mild, so the system is (approximately) optimally operated
at steady state in the sense of Angeli, Amrit, and Rawlings' Definition~6.1 of
\citep{angeli_average_2012}. In that regime the average-performance
bound of Theorem~1 is attained essentially tightly at any horizon
long enough for the terminal condition to be attainable at
negligible slack, i.e.\ long enough that the periodic orbit can be
reached from the current state within $N$ steps, which under a hard
terminal constraint is the recursive feasibility threshold and under
the soft tube used here is the point at which the slack penalty
stops being paid. Because the optimal trajectory is
already near the steady state, extra lookahead beyond that minimum
has nothing left to improve, so no lookahead elbow can exist. A pronounced elbow at the price period (here 24\,h) would
arise only if optimal operation were genuinely periodic, i.e., if
day-scale storage arbitrage were profitable, in which case the
closed loop must see a full price cycle before its cost converges to
the orbit cost. Under the present price and efficiency parameters
that arbitrage is not profitable, so the shortest tested horizon
already suffices.

Third, the terminal
penalty weight (activation and plateau values reported in
Section~\ref{sec:terminal_sweep}) controls an alignment-vs-window-cost
trade rather than a genuine cost-improving activation: the AAC shift
at activation largely reflects the storage inventory the anchored
runs are forced to preserve, and the unanchored runs deplete, rather
than a genuine economic effect of either sign. The discrete CHP
commitment schedule is identical across the entire sweep, so the
terminal weight steers the storage states without touching the
integer dispatch.

  Where inside the recommended plateau (Section~\ref{sec:grid_aac}) to sit is a
use-case trade: large weights approach a hard terminal constraint, smaller
ones leave the closed loop some deviation freedom, down to the battery 's
saturation point below which the extra freedom is bought by giving up
battery alignment outright, not by relaxing it gradually. The battery is the
component that binds: the DHN turnpike fraction has already reached its
plateau at $w_{\mathrm{term}} = 2\cdot10^2$ (mean $0.96$ across horizons on the
primary week, and it never saturates at unity anywhere in the grid), whereas
the battery SOC fraction reaches $1.000$ at every horizon only from
$3\cdot10^2$ upward. Setting $w_{\mathrm{term}} = 10^3$ therefore keeps the
battery anchored with a margin over that edge while sitting at the low end of
the cost plateau, which is why we recommend it as the default.

Fourth, the TES exhibits a persistent receding-horizon drift
that the horizon and terminal-weight sweeps together isolate as a
mechanism distinct from CHP dispatch. The TES
mechanism has four parts. (i) The TES energy level carries no direct
stage cost, only passive UA heat losses, so holding energy at any
level is economically free to first order. (ii) Each receding-horizon
solve satisfies the terminal condition by scheduling the required
discharge in the horizon tail; because the horizon shifts forward at
every re-solve, that scheduled discharge is never actually executed:
the planned terminal TES deviation is zero at every individual solve,
while the realized closed-loop level ratchets upward over time. (iii)
Sweeping $w_{\mathrm{term}}$ from $0$ to $10^{15}$
(Section~\ref{sec:terminal_sweep}) leaves the TES drift behavior
statistically unchanged, confirming that the terminal penalty is not
the driver of this drift. (iv) The offline periodic reference cannot
exhibit the same behavior because its hard periodicity constraint
$x(0) = x(N_p)$ forces intra-period charge/discharge symmetry, and
under that symmetry the lossy, power-limited TES is dominated by the
DHN's own thermal buffering, which is why the reference keeps the TES
nearly inert (Section~\ref{sec:baseline}). Theoretically, the TES
level is a direction in which the stage cost is flat, so strict
dissipativity fails along that coordinate; Theorem~1's
average-performance bound still holds and is met closely
(Section~\ref{sec:grid_aac}), but convergence to the
reference trajectory is not implied in that direction, and the
observed TES drift is exactly the kind of behavior that
performance-without-dissipativity results permit. The
economic redundancy of the TES is structural rather than an artifact
of scenario sizing. Three sensitivity experiments on the offline
periodic reference confirm this: scaling the heat demand by $1.5$
leaves the reference TES trajectory unchanged (the heat pump absorbs
the additional load within its capacity headroom); tightening the DHN
temperature band from $[90,95]$\,°C to $[92,95]$\,°C activates the
TES only marginally, at modestly higher cost; and
quadrupling the TES power limit from $0.5$\,MW to $2$\,MW leaves the
dispatch and cost unchanged. In each case the thermal inertia of the
district-heating water mass provides the buffering the lossy,
power-limited TES would otherwise supply.
The principal limitation of this study is the assumption of perfect foresight. In practice, forecast errors in load, price, and renewable generation will degrade alignment between online trajectories and the periodic reference. A second, related limitation is that the closed loop is simulated with the plant model identical to the controller's own internal model: there is no plant--model mismatch in these experiments, so all reported results are a nominal self-consistency demonstration rather than a test of robustness to modeling error. The soft terminal mechanism provides inherent robustness by permitting controlled deviations when forecasts are inconsistent with the offline orbit, but a systematic evaluation under uncertainty and under plant--model mismatch is left for future work.

\section{Conclusion}
\label{sec:conclusion}

This paper presented a cyclic-terminal EMPC framework for
coordinated operation of integrated thermal and electrical energy
networks, combining reduced-order state-space models of district
heating, storage, generation, and DC power flow within a single
mixed-integer optimization.
The key methodological contribution is the use of a periodic
economic reference trajectory computed offline and enforced online
through soft terminal constraints with scaled slack penalties. This
mechanism anchors receding-horizon optimization to an economically
meaningful periodic regime without imposing hard terminal
constraints.
Simulation results on a
campus-scale system demonstrated an asymmetric flexibility
allocation the reverse of a thermal-dominates-medium-term picture:
the battery performs slow, multi-hour price arbitrage and is the
medium-term flexibility mover, while the TES is a fast,
near-continuous cycler driven primarily by a persistent heat-pump
overproduction bias rather than by price (Section~\ref{sec:baseline}).
Under perfect foresight, the online EMPC tracked the periodic
orbit for the continuous network states, with deviations
concentrated in storage cycling and CHP commitment and the
remaining cost gap attributable to finite-horizon effects and
hybrid feasibility constraints. Horizon sensitivity analysis
(Section~\ref{sec:grid})
showed
that closed-loop performance already saturates at the shortest
tested horizon, $N=72$ steps (6\,h, half a diurnal cycle); $N=288$
nonetheless remains the practical lookahead recommendation, since it
matches $N=72$ on cost while retaining a low-switching discrete
schedule and spanning a full diurnal cycle, against a mean solve
time that grows $\approx 12.3\times$ from $N=72$ to $N=864$. The
recommended terminal weight is $w_{\mathrm{term}} = 10^3$, the
baseline choice used throughout, with any value in $[10^3, 10^{12}]$
equivalent in closed loop; within this
range the terminal tube is active for all states jointly, AAC is flat, and
battery, DHN, HP, and CHP all track the reference closely. These
lookahead and weight recommendations are established for this
campus system and the studied price weeks; we have not tested
whether they generalize to other systems or price regimes.
Beyond these design recommendations, the analysis identified a
structural property of receding-horizon EMPC applied to hybrid
systems that is not about CHP commitment but about thermal storage:
the online EMPC actively drives the TES
far across its range while the offline periodic reference, bound
by its hard periodicity constraint, keeps it nearly inert. This
receding-horizon TES drift persists across the full terminal-weight
sweep and is best understood as a direction in which the stage cost
is flat, so that Theorem~1's average-performance bound is met
closely without implying trajectory convergence in that
coordinate.
Future work will address forecast uncertainty and model mismatch
through robust and stochastic extensions of the cyclic-terminal
formulation, temperature-dependent heat-pump performance, and
higher-fidelity thermal network models. In particular, we intend to
treat forecast uncertainty in load, price, and renewable generation
via distributionally robust optimization, using the Wasserstein-metric
framework of \mbox{
\citep{mohajerin_esfahani_data-driven_2018, recke_distributionally_2026} }\hskip0pt
to construct
ambiguity sets directly from historical forecast-error data rather
than assuming a fixed distribution or bounded-support scenario set.
A quantitative comparison against established MILP-based
scheduling approaches, such as day-ahead unit commitment,
will serve to benchmark the closed-loop economic performance under
realistic operational conditions. The TES receding-horizon
drift identified here motivates extensions that restore strict
dissipativity in the storage-level coordinate. Theorem~3 of
\citep{angeli_average_2012} provides the constructive route:
augmenting the stage cost with a convex regularization term on the
storage level restores strict dissipativity in that coordinate and
thereby recovers closed-loop convergence to the reference orbit, at
a quantifiable cost overhead; an ablation quantifying this trade-off
is left for future work.

\section*{CRediT authorship contribution statement}
\textbf{Abdul Azzam:} Conceptualization, Data curation, Formal analysis, Investigation, Methodology, Project administration, Resources, Software, Supervision, Validation, Visualization, Writing – original draft, Writing – review \& editing. \textbf{Lukas Schwenkel:} Formal analysis, Investigation, Methodology, Validation, Writing – review \& editing. \textbf{Leon Scheurer:} Formal analysis, Investigation, Methodology, Software, Writing – original draft. \textbf{Pascal Häbig:} Validation, Supervision, Writing – original draft. \textbf{Kai Hufendiek:} Funding acquisition, Writing – review \& editing.
\section*{Declaration of competing interest}
The authors declare that they have no known competing financial interests or personal relationships that could have appeared to influence the work reported in this paper.
\section*{Acknowledgements}
This work was supported by the Carl Zeiss Foundation within the Stuttgart Research Initiative ``Discursive Transformation of Energy Systems'' (SRI DiTEnS).
\section*{Data and Code Availability}
The data and code that support the findings of this study are available from the corresponding author upon reasonable request.
\section*{Declaration of generative AI and AI-assisted technologies in the manuscript preparation process}
The authors used Claude (Anthropic) to support code development (Python simulation and analysis scripts), LaTeX formatting, and manuscript text editing. All code was reviewed for correctness and validated against simulation results; all text was edited for accuracy and consistency. The authors take full responsibility for the content of the published article.

\bibliographystyle{elsarticle-num}
\bibliography{doc/SESAAU}

\end{document}